\documentclass[aps,prd,superscriptaddress,twocolumn,nofootinbib,floatfix,notitlepage]{revtex4-1}

\usepackage{graphicx}
\usepackage{amsmath,amsfonts,amssymb}
\usepackage{hyperref}
\usepackage{color}
\usepackage{verbatim}

\newcommand{\be}{\begin{equation}}
\newcommand{\ee}{\end{equation}}
\newcommand{\bea}{\begin{equation}\begin{aligned}}
\newcommand{\eea}{\end{aligned}\end{equation}}
\newcommand{\td}{{\rm d}}
  
\newcommand{\vev}[1]{\langle #1 \rangle}  

\begin{document}

\title{Escape from supercooling with or without bubbles: gravitational wave signatures}

\author{Marek Lewicki}
\email{marek.lewicki@fuw.edu.pl}
\affiliation{Faculty of Physics, University of Warsaw ul.\ Pasteura 5, 02-093 Warsaw, Poland}

\author{Oriol Pujol\`as}
\email{pujolas@ifae.es}
\affiliation{Institut de Fisica d'Altes Energies (IFAE), The Barcelona Institute of Science and Technology, Campus UAB, 08193 Bellaterra (Barcelona), Spain}

\author{Ville Vaskonen}
\email{vvaskonen@ifae.es}
\affiliation{Institut de Fisica d'Altes Energies (IFAE), The Barcelona Institute of Science and Technology, Campus UAB, 08193 Bellaterra (Barcelona), Spain}

\begin{abstract}
Quasi-conformal models are an appealing scenario that can offer naturally a strongly supercooled phase transition and a period of thermal inflation in the early Universe. A crucial aspect for the viability of these models is how the Universe escapes from the supercooled state. One possibility is that thermal inflation phase ends by nucleation and percolation of true vacuum bubbles. This route is not, however, always efficient. In such case another escape mechanism, based on the growth of quantum fluctuations of the scalar field that eventually destabilize the false vacuum, becomes relevant. We study both of these cases in detail in a simple yet representative model. We determine the duration of the thermal inflation, the curvature power spectrum generated for the scales that exit horizon during the thermal inflation, and the stochastic gravitational wave background from the phase transition. We show that these gravitational waves provide an observable signal from the thermal inflation in almost the entire parameter space of interest. Furthermore, the shape of the gravitational wave spectrum can be used to ascertain how the Universe escaped from supercooling.
\end{abstract}

\maketitle

\section{Introduction}

The history of the very early Universe is largely unknown~\cite{Allahverdi:2020bys}. The earliest epoch whose existence is supported by observations is cosmic inflation. The cosmic microwave background (CMB) observations~\cite{Akrami:2018odb} limit the energy scale of inflation to be below $\mathcal{O}(10^{16} \,{\rm GeV})$ and the amplitude of curvature perturbations generated during inflation on length scales larger than $\mathcal{O}({\rm Mpc})$ to be small, $\mathcal{O}(10^{-9})$. Moreover, the cosmic inflation has to be long enough in order for the CMB scales to exit horizon. Inflation can, however, consist of multiple separate periods, in between which the Universe was, for example, radiation dominated. The CMB observations constrain only the energy scale of the period of inflation during which the CMB scales exited horizon for the last time, and the spectrum of the perturbations generated on those scales.

One reasonable possibility is that after the primordial inflation, during which the CMB scales exited horizon, the Universe experienced a period of thermal inflation. Such period can be naturally realized in models of spontaneous symmetry breaking driven by a scalar field $\phi$: At high temperatures the effective potential of the scalar field has only one symmetry-preserving minimum, $\vev{\phi}=0$, but as the temperature falls symmetry breaking minima can appear. The $\phi=0$ minimum may at some point become a false vacuum. If the field stays in it for a sufficiently long time, the vacuum energy of the false vacuum can eventually become the dominant energy density component. Trapped in the false vacuum, the Universe experiences a period of exponential expansion also known as thermal inflation~\cite{Lyth:1995ka,Lyth:1995hj}.\footnote{By thermal inflation we refer to the whole period of exponential expansion that started with the scalar field being trapped into the false vacuum, and may continue, as we will describe in Sec.~\ref{sec:fluctuations}, down to very small temperatures while  the scalar field evolves slowly  towards the true vacuum.} The fate of the Universe then depends on the details of the symmetry breaking dynamics, which are mainly encoded in the scalar effective potential $V(\phi)$.

Quasi-conformal models provide a particularly simple and well-motivated realization precisely of this second stage of inflation. By quasi-conformal, we refer to models featuring a nearly scale invariant potential, $V(\phi)=\lambda(\phi) \phi^4/4$, with a slowly scale-dependent effective quartic coupling $\lambda(\phi)$, which leads generically to a nontrivial conformal symmetry breaking minimum $\vev\phi\neq0$. Typically this transition can be very strongly supercooled and thus a long period of thermal inflation can be realized in these models. 

The epoch of thermal inflation may end in two very distinct ways in these scenarios. One possibility is by nucleation of bubbles of the true vacuum. This option has been extensively studied in various contexts. It is possible, however, that the bubble nucleation is too slow to be able to finish the transition.\footnote{In models where the order parameter $\phi$ couples to QCD or to the Higgs, QCD effects can play an important role and give a chance to transition to the true vacuum~\cite{Witten:1980ez,Quigg:2009xr,Iso:2017uuu,Hambye:2018qjv,vonHarling:2017yew,Marzo:2018nov,Baratella:2018pxi}.} In this case, at first sight it might seem that the thermal inflation phase would continue forever. However, in quasi-conformal models this is not the case. The potential energy barrier between the true and false vacua arises purely by the thermal effects, the potential around the false vacuum eventually becomes very flat and the quantum fluctuations of the scalar field are enough to escape from the false vacuum.

In this paper we consider both of the above mechanisms for exiting thermal inflation, paying particular attention to the latter, that is less discussed in literature, in the context of quasi-conformal models. We re-derive numerically early results by Vilenkin~\cite{Vilenkin:1982wt,Vilenkin:1982sg,Vilenkin:1983xp} for the duration of thermal inflation, and go further by calculating the curvature power spectrum generated for the scales that exit the horizon during this stage. Finally, we study the gravitational wave (GW) signal from the transition, that in the first case, where the transition is completed by bubble nucleation and percolation, arises from bubble collisions and motions in the plasma, and in the second case, where the fluctuations of the scalar field destabilize the false vacuum, is sourced by the scalar fluctuations. In both cases we find that the GW signal is strong and allows almost entirety of the relevant parameter space to be probed with future GW detectors.

\section{Thermal inflation from quasi-conformal dynamics}
\label{sec:inflation}

There are two types of well-understood quasi-conformal models, whose scalar potential takes the form $ \lambda(\phi) \phi^4/4 $ with slowly evolving $\lambda(\phi)$. The first type describes a confinement-deconfinement transition in (nearly conformal) strongly coupled gauge theories, and it has been modelled mainly using AdS/CFT methods~\cite{Creminelli:2001th} (see also~\cite{Randall:2006py,Konstandin:2011ds,Nardini:2007me,Konstandin:2011dr,Iso:2017uuu,Bunk:2017fic,vonHarling:2017yew,Dillon:2017ctw,Megias:2018sxv,Bruggisser:2018mrt,Baratella:2018pxi,Aoki:2019mlt,vonHarling:2019gme,DelleRose:2019pgi,Azatov:2020nbe}). The second class of models consists in the radiative symmetry breaking in a weakly coupled gauge theory {\`a} la Coleman-Weinberg~\cite{Coleman:1973jx} (see also~\cite{Jinno:2016knw,Jaeckel:2016jlh,Marzola:2017jzl,Prokopec:2018tnq,Hambye:2018qjv,Marzo:2018nov,Baldes:2018emh,Fujikura:2019oyi,Wang:2020jrd,Ellis:2020nnr}). We are going to consider that the quasi-conformal dynamics takes place in a hidden sector, and so the physical distinction between the two cases is that the hidden sector is a confining or a higgsing gauge theory. In the near-conformal limit, both types of phenomena proceed by similar phase transitions where the initial vacuum (deconfined or unbroken) becomes a supercooled false vacuum at low temperature, with the height of the potential barrier controlled by thermal effects. In the following we will use the radiative breaking of a $U(1)$ gauge theory, as a simple yet representative example. We expect our most important results to be exportable to other cases.  

More specifically, in addition to the Standard Model fields, we assume the existence of a hidden Abelian gauge field and a complex scalar $\Phi$. We start from the classically scale invariant scalar potential for the modulus of the scalar $\phi=|\Phi|$, $V(\phi) = \lambda \phi^4/4$. The radiative zero temperature corrections to such scalar potential reveal a symmetry breaking minimum~\cite{Coleman:1973jx}, and the thermal corrections give raise to a potential energy barrier that separates that minimum from the symmetric one at $\phi=0$. For example, in models similar to the classically scale invariant scalar electrodynamics, at one loop in flat space the effective potential in the high temperature approximation is given by\footnote{We note that while the high-$T$ approximation qualitatively captures the transition dynamics, for accurate mapping between the model parameters and the characteristics of the transition the full one loop effective potential with thermal resummation should be used.}
\be \label{eq:pot}
V(\phi) = V_0 + \frac{3 g^4}{4\pi^2} \phi^4 \left[\ln \frac{\phi^2}{v^2} - \frac12 \right] + \frac{g^2 T^2}{2} \phi^2 \,,
\ee 
where $V_0$ denotes the vacuum energy at $\phi=0$, $v$ the vacuum expectation value of $\phi$, $T$ the plasma temperature, and $g$ the $U(1)$ gauge coupling constant.\footnote{
Using a physical ultraviolet cutoff $\Lambda$ to regulate the loop integrals, the gauge boson induces cutoff-sensitive mass corrections to the scalar, $\delta m^2\sim g^2 \Lambda^2$. Technically, it is still possible to maintain quasi-conformality (only logarithmic deviations from scale invariance) by adjusting a mass counter-term, even though this would be seen as a fine tuning. Note that in quasi-conformal models of confining type do not present this issue.} By requiring that in the true minimum the vacuum energy vanishes we find $V_0=3 g^4 v^4/(8\pi^2)$. For simplicity, we consider minimal coupling to gravity.

The symmetry breaking vacuum becomes degenerate with the symmetric vacuum at $T\approx 0.3 g v \equiv T_c$, and it is not uncommon for the transition to occur so late that the vacuum energy $V_0$ starts to dominate over the radiation energy density $\rho_\gamma = \pi^2 g_* T^4/30$. Assuming $g_*\approx 100$ for the effective number of relativistic degrees of freedom, the vacuum energy dominance starts when the temperature drops below $T \approx 0.16 g v \equiv T_v$, and at $T\ll T_v$ Hubble rate is constant,
\be
H \approx \sqrt{\frac{8\pi V_0}{3M_P^2}} = \frac{g^2 v^2}{\sqrt{\pi} M_P} \,, 
\ee
where $M_P$ denotes the Planck mass. The Universe then experiences a period of thermal inflation.

As we will show in Sec.~\ref{sec:Pzeta}, the curvature power spectrum generated for the scales that exit the horizon during thermal inflation is very different from the one observed from the CMB. We therefore need to assure that the CMB scales do not exit the horizon during the thermal inflation stage. The comoving wavenumber (with $a=1$ today) of the smallest scales for which the Planck CMB observations put stringent constraints is $k_{\rm max} \approx 0.5\,{\rm Mpc}^{-1}$~\cite{Akrami:2018odb}. Assuming instantaneous reheating to temperature $T_{\rm reh}$ with effective number of relativistic degrees of freedom $g_*\approx 100$, and standard radiation dominated early universe expansion history, these scales exited horizon 
\be \label{eq:Nmax}
N_{\rm max} = 23.8 + \ln\frac{T_{\rm reh}}{\rm TeV}
\ee
$e$-folds before the end of inflation. If there was a period of thermal inflation following the primordial one, Eq.~\eqref{eq:Nmax} gives an upper bound on its length, $N_{\rm th} < N_{\rm max}$. Saturating the upper bound on the scale of inflation set by the Planck observations, $H<2.5\times 10^{-5} M_P$~\cite{Akrami:2018odb}, which implies $T_{\rm reh} < 6.6\times 10^{15}\,{\rm GeV}$, we get 
\be
N_{\rm th} < 53.3 \,.
\ee

It is worth noticing that a long period of thermal inflation affects the modelling of the primordial inflation. For example, in Starobinsky-like inflation one finds the spectral index to be $n_s=1-2/N_{\rm pr}$, where $N_{\rm pr}$ denotes the number of $e$-folds of primordial inflation realised after the Planck pivot scale, $k_* = 0.002\,{\rm Mpc}^{-1}$, exited the horizon. Given that the Planck measurements set the spectral index to be very close to unity, $n_s=0.965 \pm 0.004$~\cite{Aghanim:2018eyx}, a long period of primordial inflation $N_{\rm pr}\approx 57$ would be preferred. In the scenario with a period of thermal inflation the CMB scales exit horizon closer to the end of the primordial inflation by the number of $e$-folds of inflation realised by thermal inflation, $N_{\rm th}$. Next we will study the length of thermal inflation in the model introduced above.

\section{Bubble Nucleation}
\label{sec:nucl}

In this section we consider the case where the nucleation of bubbles of the true vacuum ends the thermal inflation stage. The near scale invariance of $V(\phi)$ guarantees that the dominant decay mode is by thermal bounce with $O(3)$ symmetry~\cite{Witten:1980ez}. We compute the false vacuum decay rate in the standard way starting from the bubble nucleation rate~\cite{Linde:1981zj}
\be \label{eq:GammaBubble}
\Gamma(T) \simeq T^4\left(\frac{S_3}{2\pi T}\right)^{\!\frac{3}{2}} e^{-\frac{S_3}{T}}\,,
\ee
where the action of the field in an $O(3)$ symmetric configuration reads
\be
S_3 = 4\pi \int r^2 {\rm d}r \left[ \frac{1}{2}\left(\frac{{\rm d}\phi}{{\rm d}r}\right)^2 + V_{\rm eff}(\phi,T) \right] \,,
\ee
Consequently, the corresponding equation of motion takes the form
\be
\frac{{\rm d}^2 \phi}{{\rm d} r^2} + \frac{2}{r} \frac{{\rm d} \phi}{{\rm d} r} = \frac{{\rm d} V_{\rm eff}}{{\rm d} \phi}\, ,
\ee
and should be solved with the boundary conditions $\phi\to0$ at $r\to\infty$ to describe the initial unstable vacuum background around the bubble and
${\rm d}\phi/{\rm d}r=0$ at $r=0$ to make the solution regular at the bubble centre. 

\begin{figure}
\centering
\includegraphics[width=\columnwidth]{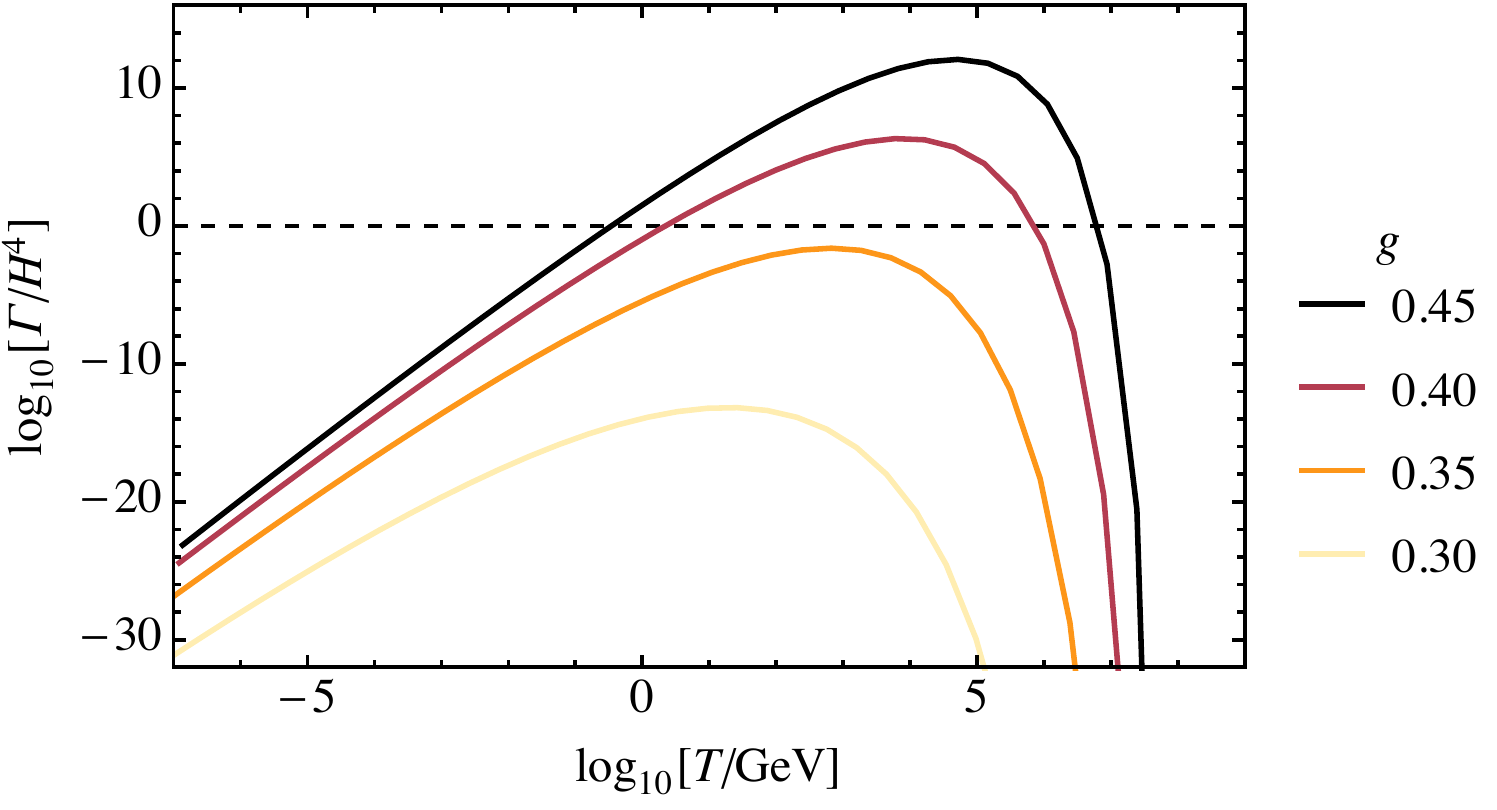}
\caption{The nucleation rate as a function of temperature for fixed $v=10^9\,{\rm GeV}$ and different values of $g$.}
\label{fig:Gamma}
\end{figure}

The probability that a given spatial point is inside a true vacuum bubble is given by $P=e^{-I(T)}$, where~\cite{Guth:1979bh,Guth:1981uk}
\be
I(T) = \frac{4\pi}{3} \int_T^{T_c} \frac{{\rm d} T' \,\Gamma(T')}{T'^4 H(T')} \left[ \int_{T}^{T'} \frac{ {\rm d} \tilde T}{H(\tilde T)} \right]^3 \,.
\ee
We estimate the bubble percolation temperature, $T_*$, by $I(T_*) = 0.34$~\cite{Ellis:2018mja}. In the class of models we consider this is well approximated by the nucleation temperature at which on average one bubble should nucleate in the horizon volume, $\Gamma \simeq H^4$. As shown in Fig.~\ref{fig:Gamma}, the bubble nucleation rate increases exponentially for decreasing $T$ near the critical temperature $T=T_c$, but at temperatures $T\ll T_c$ the action scales as $S_3\propto T$ and the false vacuum decay rate then decreases as $\Gamma \propto T^4$. It is therefore possible for the nucleation rate to always remain below the Hubble rate. In such case the bubble nucleation will not catalyse the transition.

\begin{figure}
\centering
\includegraphics[width=0.95\columnwidth]{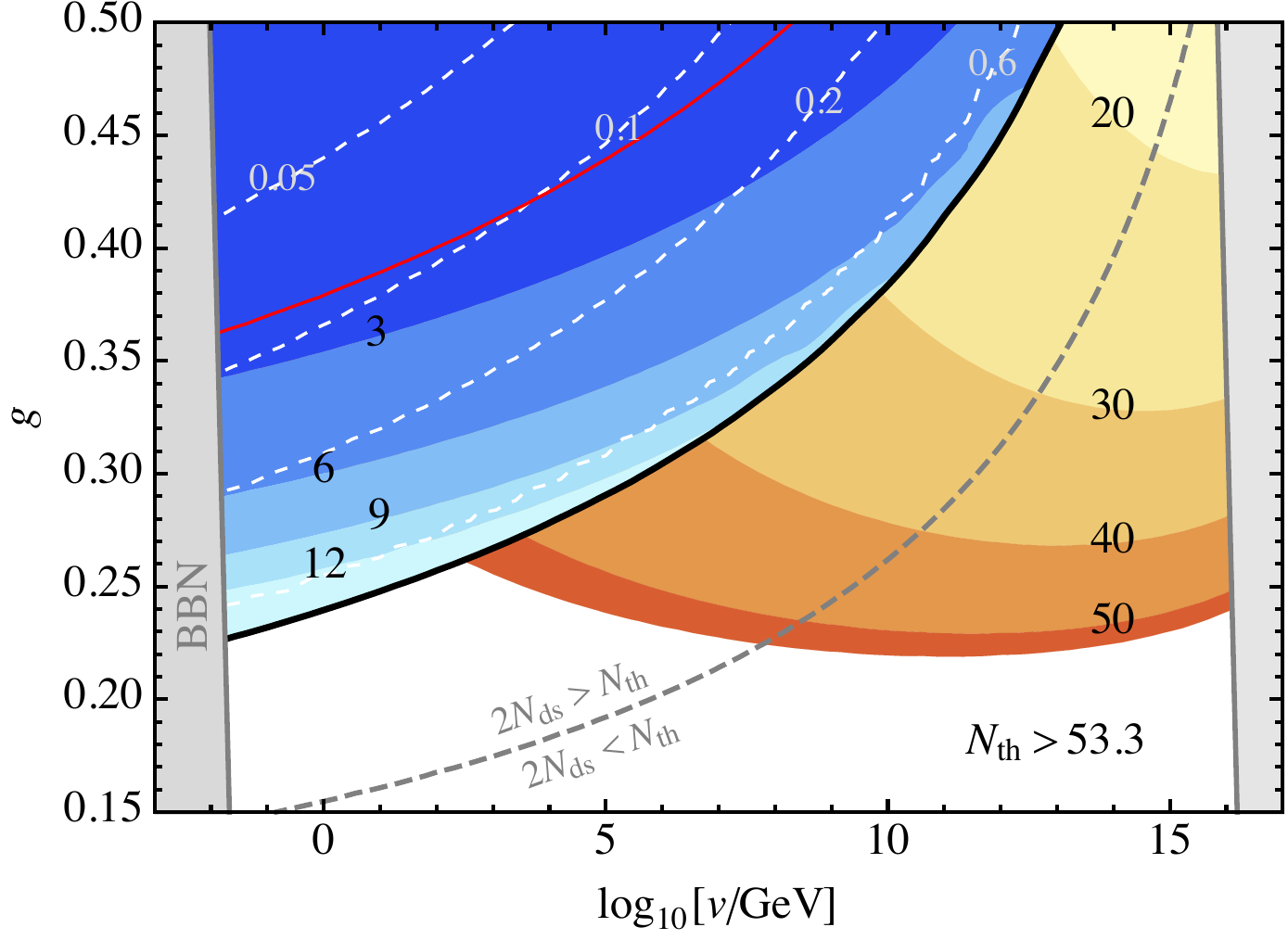}
\caption{The duration of the thermal inflation stage is shown by the color coding with the labels indicating the number of $e$-folds. Above the solid black curve the thermal inflation ends by thermal bubble nucleation, and the white dashed contours indicate the average bubble radius at percolation, $HR_*$. Below the red curve $g^2>\lambda/4$. The gray region on the left is excluded by the BBN the reheating temperature and in the gray region on the right the symmetry breaking phase transition is never realized.}
\label{fig:efolds}
\end{figure}

In Fig.~\ref{fig:efolds} the region where the bubble nucleation fails to finish the thermal inflation period corresponds to the region below the solid black curve. Above it the color coding indicates the number of $e$-folds of thermal inflation before the transition, $N = \ln T_v/T_*$. The white dashed contours instead indicate the average bubble radius at percolation,\footnote{For strong transitions, the average bubble radius at percolation is related to the often used inverse time scale $\beta$ via $\beta = (8\pi)^{1/3}/R_*$.}
\be
R_* =  \left[T_* \int_{T_*}^{T_c} \frac{{\rm d} T'}{T'^2} \frac{\Gamma(T')}{H(T')} e^{-I(T')} \right]^{-\frac13} \,,
\ee
which approaches the Hubble horizon radius $H^{-1}$ in the limit where the bubbles never percolate. In the gray region on the right the symmetry breaking phase transition is not realized as the critical temperature where the symmetric and symmetry breaking minima are degenerate, $T_c \approx 0.3 g v$, is never higher than the upper bound on the reheating temperature after primordial inflation set by the CMB observations~\cite{Akrami:2018odb}, $T_c > 6.6\times 10^{15}\,{\rm GeV}$. The gray region on the left is excluded by the BBN constraint on the reheating temperature after thermal inflation, $T_v > 5\,{\rm MeV}$~\cite{Allahverdi:2020bys}.\footnote{Throughout this paper we assume an instantanous transition to radiation dominance after thermal inflation. The details of the reheating depend strongly on the underlying model, and we leave the study of such issues for future work.}

Finally, while it might seem that gravitational effects can be important for the nucleation process given the de Sitter background, we have made sure also computing the action of the Coleman-De Luccia instantons~\cite{Coleman:1980aw} that the flat background approximation discussed above is perfectly fine in all of the parameter space of interest as the scales we discuss are significantly below the Planck scale. Therefore, the potential non-minimal coupling to gravity also plays a negligible role in the nucleation process in the models of interest~\cite{Czerwinska:2016fky}. 

\section{Growth of fluctuations}
\label{sec:fluctuations}

In the case that the bubble nucleation rate never reaches the Hubble rate, the quantum fluctuations eventually become significant and can destabilize the false vacuum. To study the growth of the fluctuations, we start from an appropriate approximation of the scalar potential. In de Sitter space, the one loop effective potential is obtained from the flat space one by taking into account the curvature induced masses ($\propto H^2$) which regulate the logarithms at the $\phi\ll H$ limit. Then, near the symmetric phase, for $v\ll M_{\rm P}$, the effective potential can be approximated as~\cite{Vilenkin:1983xp}
\be \label{eq:Vapprox}
V(\phi) \approx V_0 + \frac{m^2}{2} \phi^2 - \frac{\lambda}{4} \phi^4 \,,
\ee
where $m^2\equiv g^2 T^2$ and $\lambda \equiv 3 g^4 \ln(g^2v^2/H^2)/\pi^2$. Using the scaling of temperature as a function of the scale factor $a$, $T\propto 1/a$, we can express the effective mass of $\phi$ as $m^2 = g^2 T_v^2/a^2 \equiv M^2/a^2$, where we have chosen the beginning of thermal inflation as the reference point with $a = a_v = 1$. The constant $M$ is defined as $M = g T_v \approx 0.16 g^2 v$.

Let us express $\hat \phi$ in terms of mode functions $u_k$ as
\be \label{eq:mod}
\hat\phi(\eta,\mathbf{x}) \!= \!\int\!\! \frac{\td^{3}{k}}{\sqrt{(2\pi )^{3}a^2}}\!\left[\hat{a}_\mathbf{k}^{\phantom{\dagger}}u^{\phantom{\dagger}}_{k}(\eta)+\hat{a}_{-\mathbf{k}}^\dagger u^*_k(\eta)\right]\!e^{i\mathbf{k}\cdot\mathbf{x}},
\ee
where $\mathbf{k}$ is the co-moving momentum, $\eta$ is the conformal time, and the creation/annihilation operators are normalised as $[\hat{a}_{\mathbf{k}}^{\phantom{\dagger}},\hat{a}_{\mathbf{k}'}^\dagger]=\delta^{(3)}(\mathbf{k}-\mathbf{k}')$. The mode functions are normalized such that 
\be \label{eq:norm}
u_k(\eta) \partial_\eta u_k^*(\eta)-u_k^*(\eta) \partial_\eta u_k(\eta)=i \,.
\ee
The equation of motion of $\hat\phi$ is in the cactus approximation~\cite{Vilenkin:1983xp} given by 
\be
0 = \Box\hat{\phi} - V'(\hat\phi) \approx \left(\Box - \langle V''\rangle\right) \hat\phi \,,
\ee
where $\sqrt{-g}\,\Box = \,\partial_\mu \sqrt{-g}\, \partial^\mu$, prime denotes derivative with respect to $\hat\phi$, and  $\langle V''\rangle = m^2 - 3\lambda \langle \hat\phi^2\rangle$. The variance of $\hat\phi$ gets contributions from the vacuum and thermal fluctuations, $\langle\hat\phi^2\rangle = \langle\hat\phi^2\rangle_v + \langle\hat\phi^2\rangle_T$, which are given by\footnote{We treat renormalization in the same fashion as in Ref.~\cite{Vilenkin:1983xp} by cutting the $k$ integral for vacuum fluctuations at $aH$, that is, including only wavelengths greater than the horizon. For thermal fluctuations the relevant scale is given by the temperature and because of the thermal distribution $n_k$ a hard cut-off is not needed.}
\be
\langle \hat\phi^2 \rangle_v = \int_H^{aH} \! \frac{k^2\td k}{2\pi^2 a^2} |u_k(\eta)|^2 \,,
\ee 
and
\be
\langle \hat\phi^2 \rangle_T = \int_H^\infty \! \frac{k^2\td k}{\pi^2 a^2} n_k |u_k(\eta)|^2 \,,
\ee 
where $n_k^{-1} = e^{\sqrt{k^2+m^2}/T_v}-1$. In a de Sitter universe, $a = -1/(\eta H)$, the equation of motion implies
\be\label{eq:medS}
\ddot u_k(\eta)+\left[\omega^2 - \frac{2}{\eta^2} - 3 \lambda a^2 \langle \hat\phi^2\rangle \right] u_k(\eta)=0 \,,
\ee 
where $\omega^2 \equiv k^2 + M^2$ and dot denotes derivative with respect to $\eta$. 

\begin{figure}
\centering
\includegraphics[width=0.95\columnwidth]{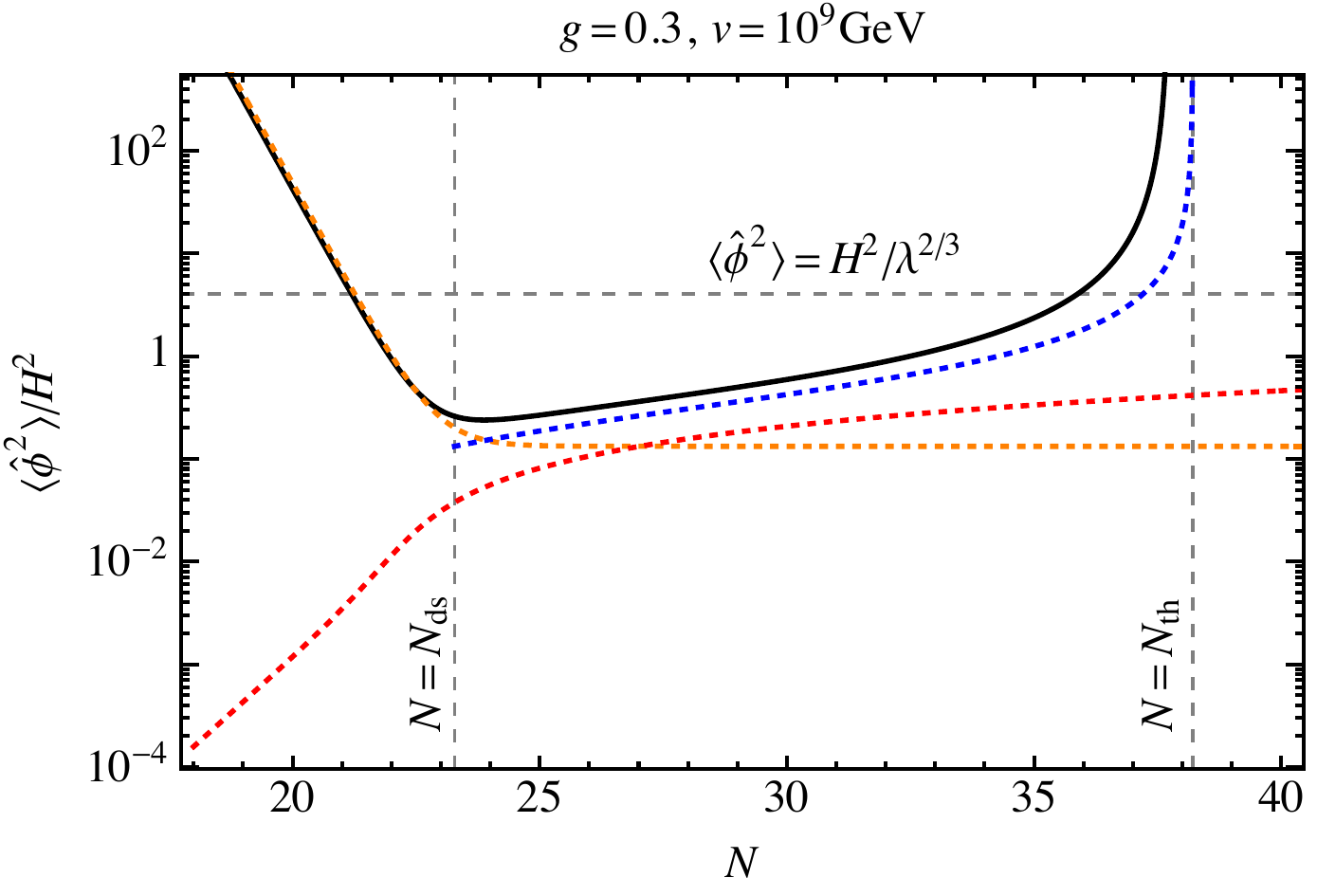}
\caption{Evolution of the variance of the fluctuations of the scalar field as a function of $e$-folds of inflation. The solid black curve shows the numerical result, the red and orange dashed curves the vacuum and thermal contributions obtained from the BD solution, and the blue dashed curve the tangent approximation~\eqref{eq:vartan}. The vertical dashed lines indicate the moment when the symmetric phase becomes unstable, $N=N_{\rm ds}$, and the end of thermal inflation, $N=N_{\rm th}$. The horizontal dashed line marks the transition to the classical regime.}
\label{fig:variance}
\end{figure}

We use the normalization condition~\eqref{eq:norm} to eliminate the phase of $u$ from the mode equation~\eqref{eq:medS} and then solve it for the amplitude $|u_k(\eta)|$. In Fig.~\ref{fig:variance} the solid black curve shows the variance $\langle\hat\phi\rangle$ as a number of $e$-folds, $a=e^N$, obtained by numerically solving the mode equation for $g=0.3$ and $v=10^9\,{\rm GeV}$. The red, orange and blue dashed curves instead correspond to analytical approximations that we will discuss below. We see that these approximations describe well the evolution of the fluctuations.

At the early stages of thermal inflation, when the false vacuum is still stable, $\langle V'' \rangle >0$\,, the last term in the mode equation~\eqref{eq:medS} can be neglected. In that regime the solution that is properly normalised is given in terms of the Hankel functions as
\be
u_{k}(\eta) = \sqrt{\frac{-\pi\eta}{4}} \left[ c_1 H_{3/2}^{(1)}(-\omega\eta) + c_2 H_{3/2}^{(2)}(-\omega\eta) \right],
\ee 
with $c_1^2 + c_2^2 = 1$. As shown in~\cite{Vilenkin:1982wt} the solution in de Sitter quickly approaches the Bunch-Davies (BD) solution, $c_1 = 0$, $c_2 =1$~\cite{Bunch:1978yq}. In Fig.~\ref{fig:variance} the thermal and vacuum contributions to the variance in this approximation are shown by the orange and red dashed curves. The vacuum part is given by~\cite{Vilenkin:1982sg,Linde:1982uu} $\langle\hat\phi^2\rangle_v \approx H^2 N/(4\pi^2) + C H^2$, where $C\sim 1$, and the thermal part by~\cite{Starobinsky:1982ee,Vilenkin:1983xp} $\langle\hat\phi^2\rangle_T \approx T^2/12 + H^2/(8\pi g)$.

The variance grows as the potential at $\phi\approx 0$ becomes flatter for decreasing $T$. Eventually, the symmetric phase becomes unstable as $\langle V'' \rangle$ crosses zero~\cite{Vilenkin:1982wt}. This moment is indicated by the first vertical dashed line in Fig.~\ref{fig:variance}. At that moment the thermal fluctuations still give the dominant contribution to $\langle \hat\phi^2\rangle$. We can therefore estimate the temperature $T_{\rm ds}$ at which $\langle V'' \rangle = 0$ by taking $\langle\hat\phi^2\rangle \approx \langle\hat\phi^2\rangle_T$. This gives
\be
T_{\rm ds} \approx H \sqrt{\frac{3\lambda}{2\pi g(4g^2-\lambda)}} \,.
\ee
Note that $g^2>\lambda/4$ for the parameters for which fluctuations are relevant, as indicated by the red curve in Fig.~\ref{fig:efolds}. 

Inflation still continues after the symmetric phase has become unstable. For $\langle V'' \rangle \ll 0$ the evolution of the variance $\langle \hat\phi^2\rangle$ is dominated by the last term in the mode equation~\eqref{eq:medS}. As shown in Ref.~\cite{Vilenkin:1983xp}, the solution can be approximated in the limit where $\langle\hat\phi^2\rangle \ll H^2/\lambda$ and $(\td/\td \eta) \langle\hat\phi^2\rangle \ll a H^2 \langle\hat\phi^2\rangle$ by
\be \label{eq:vartan}
\langle\hat\phi^2\rangle \!=\! \frac{H^2}{2\pi \sqrt{2\lambda}} \,\tan\! \left[\sqrt{\frac{\lambda}{2}}\frac{N-N_{\rm ds}}{\pi} + \tan^{-1} \!\sqrt{\frac{\lambda}{8 g^2}} \right] ,
\ee
where we approximated the variance at $N = N_{\rm ds}$ as $\langle \hat\phi^2\rangle \approx H^2/(8\pi g)$\,.\footnote{In Ref.~\cite{Vilenkin:1983xp} initial condition $\langle \hat\phi^2\rangle = 0$ was used. We find that this would overestimate the duration of inflation after the symmetric vacuum has become unstable.} In Fig.~\ref{fig:variance} this approximation is shown by the blue dashed curve. Of course, before the fluctuations grow too large, one expects that this description ceases to hold. Indeed, at some point the characteristic amplitude of the field, $\phi_c \equiv \sqrt{\langle \phi^2 \rangle}$ is large enough that the classical dynamics can dominate again. This occurs when $H^3 < V'(\phi_c)$, which gives 
\be
\langle \phi^2 \rangle \sim H^2/\lambda^{2/3} \,,
\ee
indicated in Fig.~\ref{fig:variance}. This marks the end of the quantum regime in the cactus approximation, and after that the field quickly rolls to the true vacuum ending inflation. The duration of thermal inflation, $N_{\rm th}$, can then be approximated by finding when the approximation~\eqref{eq:vartan} diverges. This gives, 
\be \label{eq:Nend}
N_{\rm th} \approx N_{\rm ds} +  \frac{\pi^2}{\sqrt{2\lambda}} - \sqrt{\frac{2}{\lambda}} \pi \,\tan^{-1} \sqrt{\frac{\lambda}{8 g^2}} \,,
\ee
where 
\be
N_{\rm ds} = \ln\frac{T_v}{T_{\rm ds}} \approx \frac12 \ln \left[\frac{0.18 g^3(4g^2-\lambda)}{\pi \lambda} \frac{v^2}{H^2}\right] 
\ee
corresponds to the moment when $\langle V''\rangle = 0$. These moments are indicated by the vertical dashed lines in Fig.~\ref{fig:variance}.

In Fig.~\ref{fig:efolds} we show the duration of thermal inflation in $e$-folds. In the region above the solid black curve thermal bubble nucleation ends the thermal inflation stage. In this case thermal inflation lasts for much less than $N_{\rm ds}$ $e$-folds and therefore the effects of the fluctuations discussed above are negligible. Below the solid black curve the bubble nucleation rate never reaches the Hubble rate. In this region the labelled contours indicate $N_{\rm th}$ given in Eq.~\eqref{eq:Nend}. In the white region at the bottom $N_{\rm th} > 53.3$, and, as discussed in Sec.~\ref{sec:inflation}, thermal inflation would modify the curvature power spectrum at scales that are probed by the Planck CMB observations. Along the gray dashed curve the periods of inflation before and after the destabilization of the symmetric phase are equally long.

\section{Curvature power spectrum}
\label{sec:Pzeta}

Next we calculate the curvature power spectrum following a similar calculation done in Ref.~\cite{Dimopoulos:2019wew}. Neglecting the kinetic term, the curvature perturbation is 
\be
\zeta = \frac{H}{\dot{\rho}_\phi}\delta{\rho}_\phi \approx H\frac{V-\langle V\rangle}{\partial_t\langle V\rangle} \approx \frac{\langle V''\rangle}{\eta \partial_\eta[\langle V''\rangle\langle\hat\phi^2\rangle]} \,\delta\phi^2 \,,
\ee
where $\delta \phi^2 = \phi^2 - \langle \hat\phi^2\rangle$, and in the last step we approximated $V \approx \langle V''\rangle \phi^2/2$. Then, the curvature power spectrum is
\bea \label{eq:Pzeta}
{\cal P}_\zeta(k) \!=\! \left[\frac{\zeta}{\delta\phi^2}\right]^2\! \frac{k^3}{\pi^2} \!\int \!&\frac{\td^3 p}{(2\pi)^3 a^4} |u_{p}(\eta)|^2|u_{|\mathbf{k}-\mathbf{p}|}(\eta)|^2 \\ &\times (1+2n_p)(1+2n_{|\mathbf{k}-\mathbf{p}|})  \,,
\eea
where we restrict the momentum integral to the range from $H$ to $a H$. We evaluate the curvature power spectrum at the moment when $\langle\hat\phi^2\rangle = H^2/\lambda^{2/3}$, after which the classical dynamics takes over.\footnote{We have checked that for the scales that are already well outside horizon the curvature power spectrum remains roughly constant until that moment.} The scales that exit the horizon  after that are in the classical regime and so we simply cut the power spectrum at that scale.

\begin{figure}
\centering
\includegraphics[width=\columnwidth]{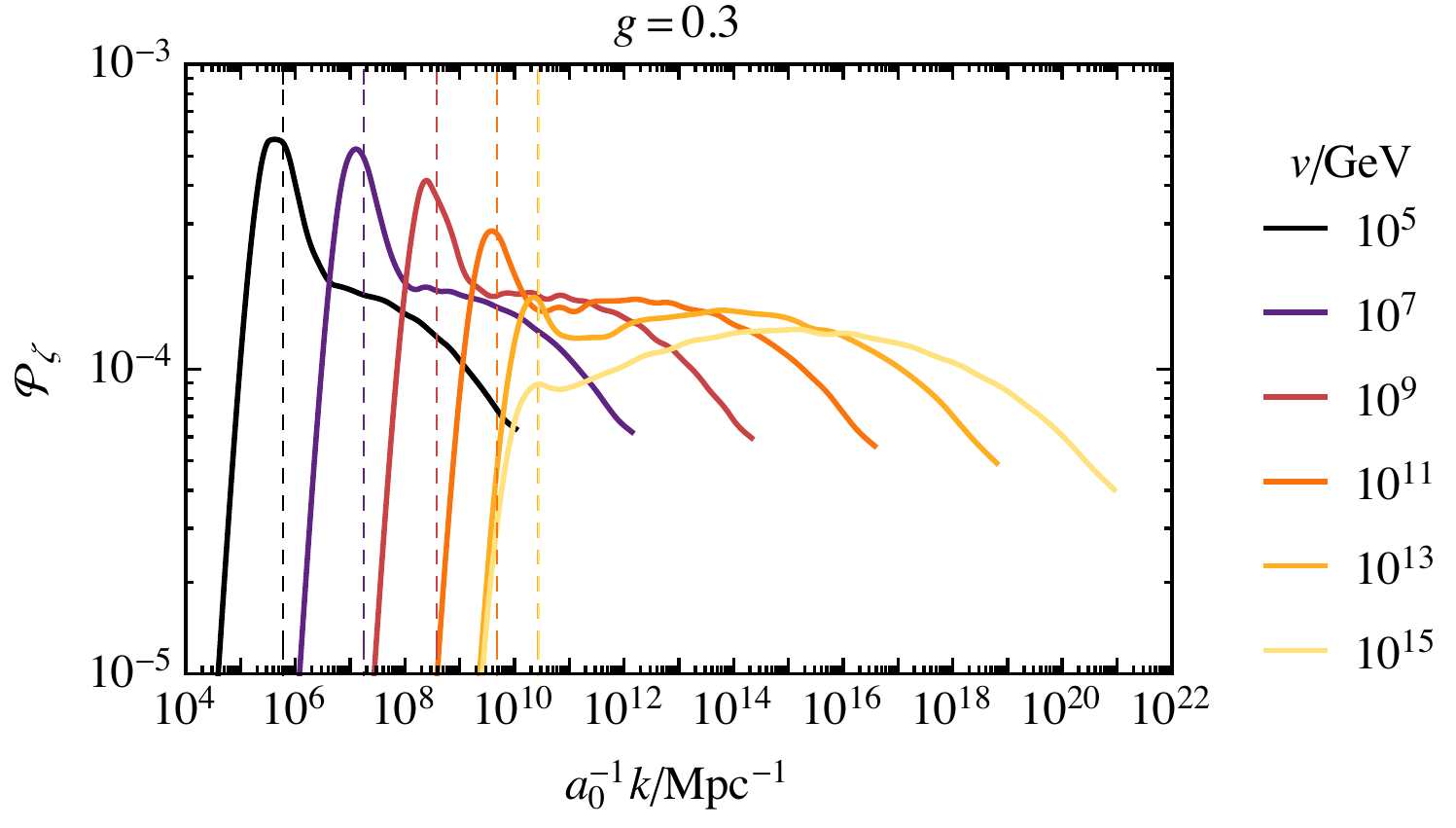} \\ \vspace{3mm}
\includegraphics[width=\columnwidth]{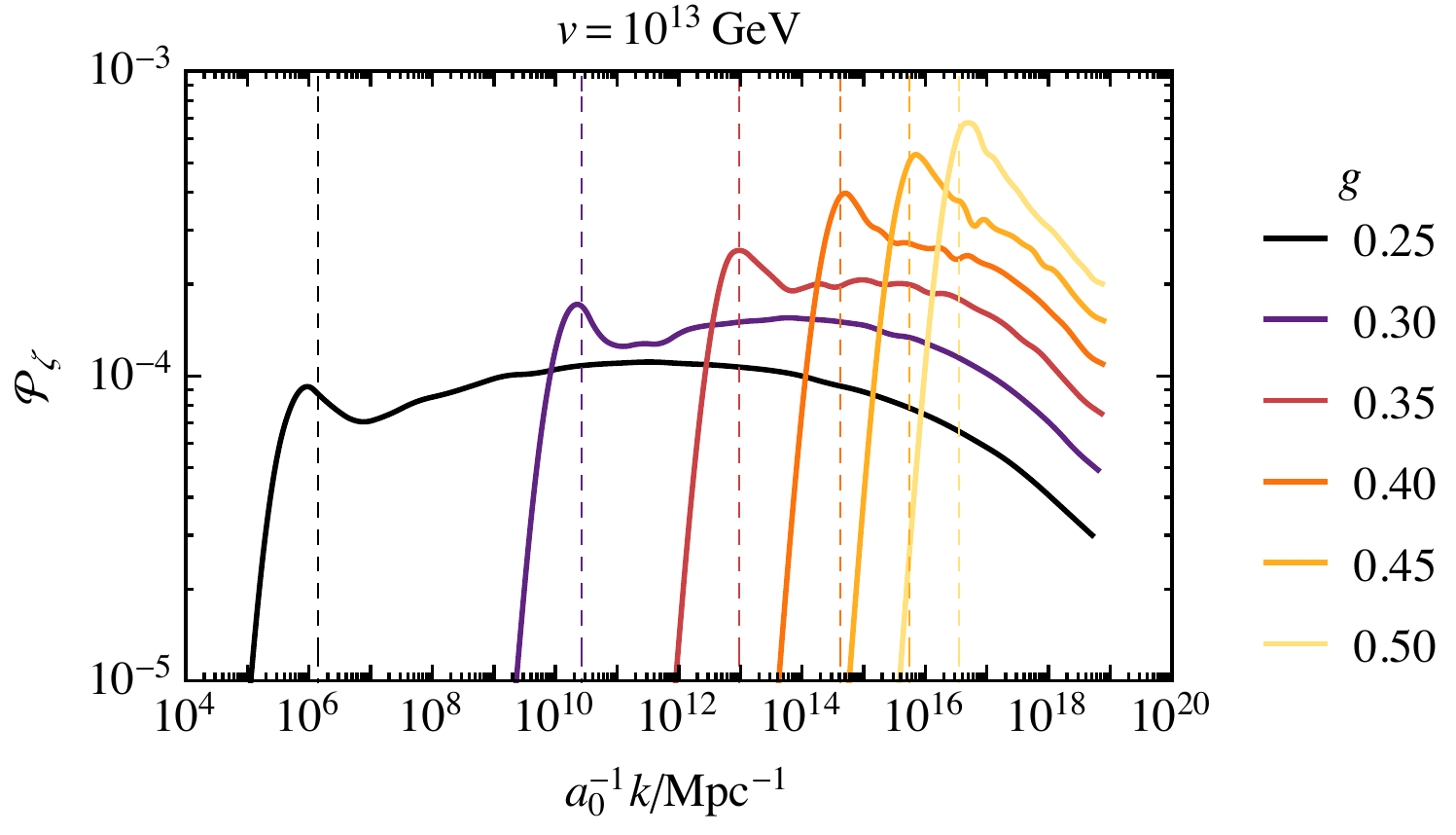}
\caption{The solid curves show the curvature power spectrum generated during the stage of thermal inflation  and the vertical dashed lines indicate the scales $k=k_{\rm ds}$ that exit the horizon when the false vacuum becomes unstable.}
\label{fig:Pzeta}
\end{figure}

In Fig.~\ref{fig:Pzeta} we show by the solid curves the curvature power spectrum integrated from the numerical solution of the mode equation~\eqref{eq:mod}. In the horizontal axis the comoving wavenumber is divided by the present scale factor, which in our normalization is 
\bea
a_0 &= e^{N_{\rm th}} \left[\frac{s(T_v)}{s(T_0)}\right]^{\frac13} \\ 
&\approx 2.1\times 10^{12} \frac{gv}{\rm GeV} \,e^{N_{\rm th}} \left(\frac{g_{*s}}{100}\right)^{\frac13} ,
\eea
where $g_{*s}$ denotes the effective number of relativistic entropy degrees of freedom at the temperature $T_v$. The spectrum grows as $\mathcal{P}_\zeta \propto k^3$ at scales that exited horizon before the symmetric phase became unstable, $N<N_{\rm ds}$. The $k^3$ growth terminates roughly at the scales that exited at $N=N_{\rm ds}$, and the spectrum has a maximum around that scale. The corresponding comoving wavenumber is
\be
\frac{k_{\rm ds}}{a_0} = \frac{e^{N_{\rm ds}} H}{a_0} \approx\! \frac{3.5\times 10^6}{\rm Mpc} \frac{g v}{\rm GeV} e^{N_{\rm ds} - N_{\rm th}} \!\left( \frac{g_{*s}}{100} \right)^{\!-\frac13},
\ee
indicated by the vertical dashed lines in Fig.~\ref{fig:Pzeta}. At higher scales the spectrum has an almost flat part, whose length depends on the duration of inflation after the symmetric phase has become unstable. The peak is dominantly generated by thermal fluctuations and the almost flat part by vacuum fluctuations.

From Fig.~\ref{fig:Pzeta} we see that the amplitude of the spectrum at $k=k_{\rm ds}$ decreases as a function of $v$ and increases as a function of $g$. We find that for all relevant parameter values for which the transition is not finished by bubble nucleation, the amplitude of the curvature power spectrum at all scales remains smaller than $10^{-3}$. As will be discussed in detail in the next section, this amplitude is sufficiently high to induce a stochastic GW background that can be probed with various future observatories. 

The amplitude of the fluctuations is not, however, high enough for formation of primordial black holes (PBHs), which would require, assuming that the relevant scales re-enter horizon during radiation dominated era,\footnote{In matter dominance the PBH formation is more efficient as the cosmic pressure is negligible, and therefore a smaller amplitude for the curvature fluctuations is enough~\cite{Khlopov:1980mg,Harada:2017fjm}. Such matter dominated era after thermal inflation could be realized for example if the decay of the thermal inflaton to radiation is slow~\cite{Carr:2017edp}. We leave the study of such a scenario for future work.} the amplitude of the curvature power spectrum to be $\mathcal{O}(10^{-2})$~\cite{Carr:1975qj}. Compared to the thermal inflation model considered in Ref.~\cite{Dimopoulos:2019wew}, where PBH formation was found to be successful, the difference is that here we consider the near conformal case whereas in Ref.~\cite{Dimopoulos:2019wew} a negative bare mass term for the scalar field was included. That scenario, however, suffers from tuning of the quartic couling, as in order to generate a high enough amplitude for the curvature perturbations the bare mass needs to be $m^2\sim H^2$, and therefore the quartic is $\lambda \lesssim m^2/M_P^2$.

\section{Gravitational Waves}

\begin{figure*}
\centering
\includegraphics[width=\columnwidth]{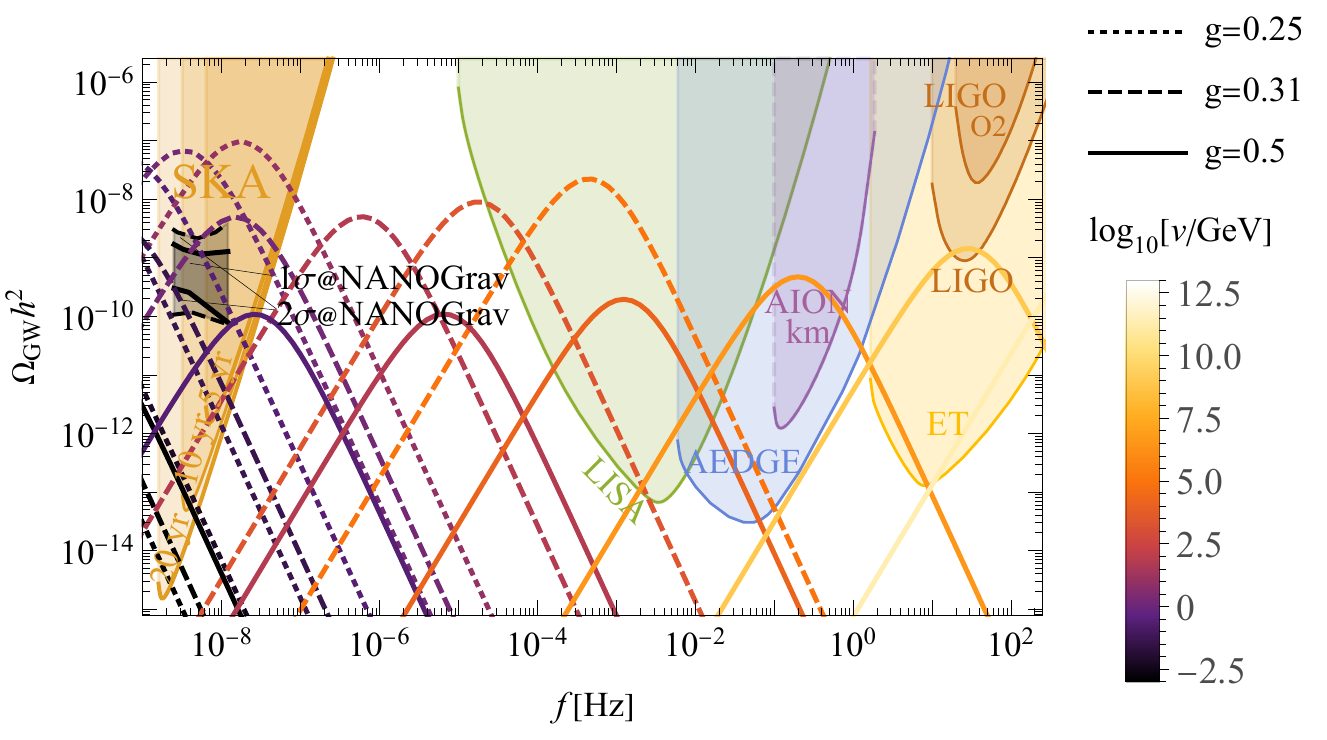} 
\includegraphics[width=\columnwidth]{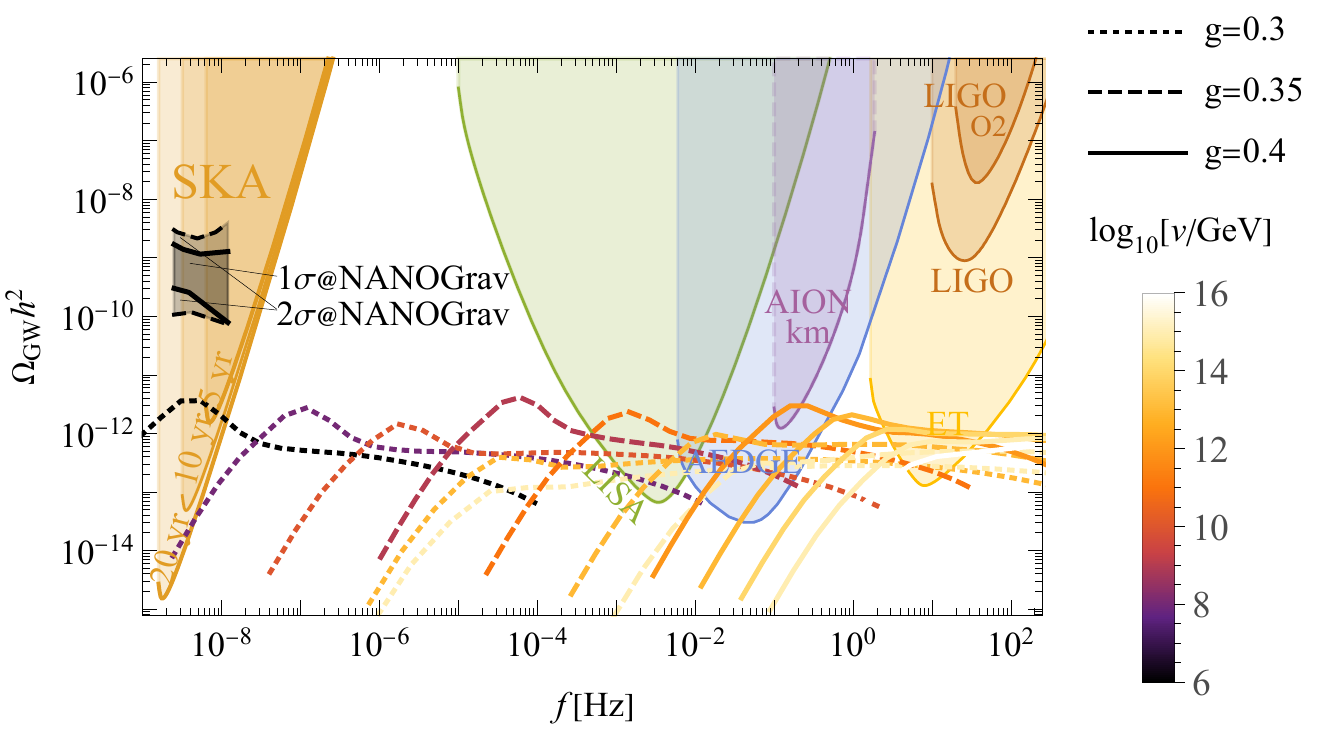}
\caption{Examples of GW spectra sourced by a bubble collisions (left panel) and the scalar field fluctuations (right panel) together with sensitivities of current and future GW detectors.}
\label{fig:GWs}
\end{figure*}

Two GW sources are relevant for the thermal inflation scenario: First, if thermal inflation finishes with nucleation of true vacuum bubbles, their collisions and the motions they induce in the plasma generate GWs. Second, in the case that the bubble nucleation can not catalyse the phase transition, the scalar field acquires large fluctuations that source GWs. We will next discuss these two cases and consider the detectability of such GWs with current and future GW observatories.

In the scenario at hand even transitions which complete are typically severely supercooled and it is reasonable to assume that the interactions of the bubble walls with the surrounding plasma are not strong enough to stop the walls from accelerating before they collide with other bubbles. In this situation the GW spectrum is dominantly sourced by the scalar field gradients~\cite{Ellis:2019oqb,Ellis:2020nnr}. In the gauge $U(1)$ case the GW spectrum sourced by the vacuum bubble collisions is given by~\cite{Lewicki:2020jiv,Lewicki:2020azd}~\footnote{In strongly supercooled transitions, $\alpha \equiv V_0/\rho_\gamma \gg 1$ and so $\alpha/(\alpha+1) \approx 1$.}
\be
\begin{split}
&\Omega_{\rm GW}^{\rm col} h^2  \approx 2\times 10^{-5} (HR_*)^2 \left(\frac{g_{*s}^4 g_*^{-3}}{100}\right)^{\!-\frac13} \\ 
&\times \left[2.94\left(\frac{f}{f_{\rm col}}\right)^{\!-0.64}+2.25\left(\frac{f}{f_{\rm col}}\right)^{\!0.84}\right]^{-3.5} ,
\end{split}
\ee
where $g_*$ and $g_{*s}$ denote the effective numbers of relativistic energy and entropy degrees of freedom~\cite{Saikawa:2018rcs} at the temperature $T_v$, and
\be
f_{\rm col} \approx 5.5\times 10^{-6}\,{\rm Hz} \,(HR_*)^{-1} \,\frac{T_v}{100\,{\rm GeV}} \left( \frac{g_{*}}{100} \right)^{\frac16}
\ee
is the peak frequency of the spectrum measured today. In the left panel of Fig.~\ref{fig:GWs} we show examples of the GW spectra from bubble collisions for various sets of parameters. We also show the power-law integrated sensitivities~\cite{Thrane:2013oya} of upcoming GW experiments LISA~\cite{Audley:2017drz}, ET~\cite{Punturo:2010zz,Hild:2010id},  AEDGE~\cite{Bertoldi:2019tck}, AION/MAGIS~\cite{Badurina:2019hst,Graham:2016plp,Graham:2017pmn},  and SKA~\cite{Janssen:2014dka} as well as already running LIGO-Virgo~\cite{TheLIGOScientific:2014jea} specifying its current sensitivity after O2 run~\cite{LIGOScientific:2019vic} which is effectively a constraint~\cite{Romero:2021kby}, and finally the spectra that can explain the recently observed stochastic common-spectrum process at NANOGrav~\cite{Arzoumanian:2020vkk}.\footnote{Various other theoretical models have been 
proposed as possible explanations of the recent NANOGrav excess~\cite{Ellis:2020ena,Vaskonen:2020lbd,DeLuca:2020agl, Kohri:2020qqd,Sugiyama:2020roc,Domenech:2020ers,Bhattacharya:2020lhc,Blasi:2020mfx, Buchmuller:2020lbh,Samanta:2020cdk,Chigusa:2020rks,Ramberg:2020oct,DeLuca:2020agl,Vaskonen:2020lbd, Kohri:2020qqd,Sugiyama:2020roc,Domenech:2020ers,Bhattacharya:2020lhc,Nakai:2020oit,Neronov:2020qrl,Vagnozzi:2020gtf, Li:2020cjj, Kuroyanagi:2020sfw, Liu:2020mru, Chiang:2020aui,Ratzinger:2020koh, Namba:2020kij,Tahara:2020fmn,Cai:2020qpu,Bian:2021lmz,Blanco-Pillado:2021ygr,Brandenburg:2021tmp,Arzoumanian:2021teu}.
}

The second relevant GW signal comes from large scalar fluctuations, that at the second order in the cosmological perturbation theory source GWs~\cite{Matarrese:1993zf,Matarrese:1997ay,Nakamura:2004rm,Ananda:2006af,Baumann:2007zm}. The spectrum of these scalar induced GWs is obtained from the curvature power spectrum $\mathcal{P}_\zeta$ as~\cite{Kohri:2018awv,Espinosa:2018eve,Inomata:2019yww}
\begin{widetext}
\be \label{Omega_GW}
\Omega_{\rm GW}^{\rm si}h^2 \approx 4.6\times 10^{-4} \left(\frac{g_{*,s}^{4}g_{*}^{-3}}{100}\right)^{\!-\frac13} \!\int_{-1}^1 {\rm d} x \int_1^\infty {\rm d} y \, \mathcal{P}_\zeta\left(\frac{y-x}{2}k\right) \mathcal{P}_\zeta\left(\frac{x+y}{2}k\right) F(x,y) \bigg|_{k = 2\pi f} \,,
\ee
where
\be
F(x,y) \!=\!\frac{(x^2\!+\!y^2\!-\!6)^2(x^2-1)^2(y^2-1)^2}{(x-y)^8(x+y)^8} \!\left\{\left[x^2-y^2+\frac{x^2\!+\!y^2\!-\!6}{2}\ln\left|\frac{y^2-3}{x^2-3}\right|\right]^{\!2} \!+\! \frac{\pi^2(x^2\!+\!y^2\!-\!6)^2}{4}\theta(y-\sqrt{3}) \right\} .
\ee
\end{widetext}
As seen from the right panel of Fig.~\ref{fig:GWs}, the shape of the scalar induced GW spectrum resembles the curvature power spectrum. In particular, the local maximum at scale $k_{\rm ds}$ in the curvature power spectrum corresponds to the frequency
\be
f_{\rm ds} = 2.0\times 10^{-5} \,{\rm Hz} \, e^{N_{\rm ds} - N_{\rm th}} \,\frac{T_v}{100\,{\rm GeV}} \left( \frac{g_{*s}}{100} \right)^{\!-\frac13}
\ee
where the GW spectrum has a maximum.

\begin{figure}
\centering
\includegraphics[width=0.95\columnwidth]{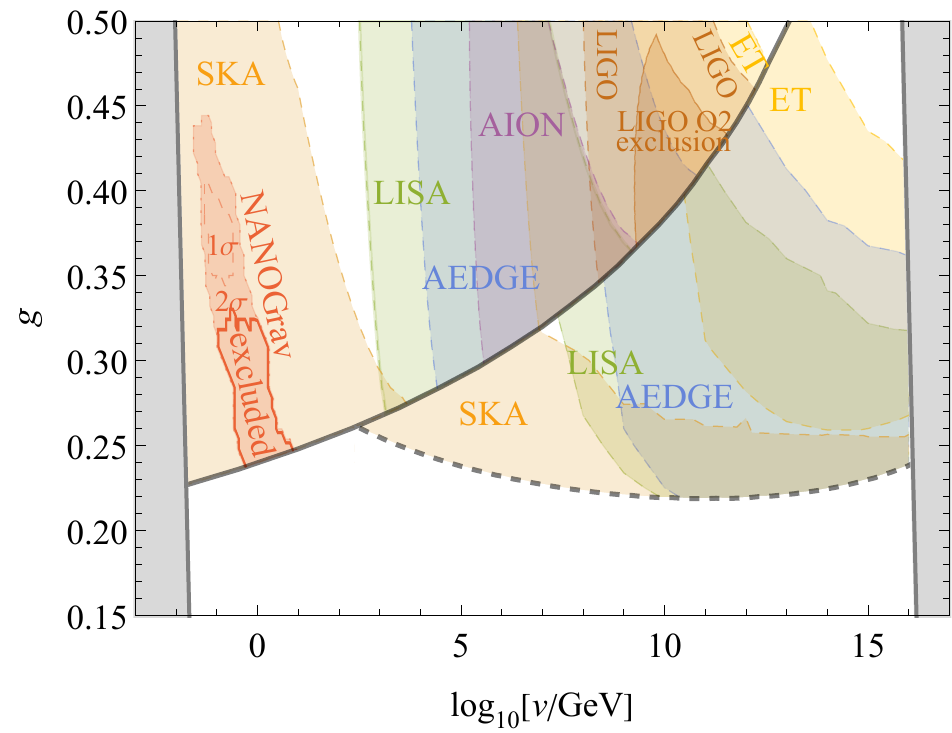}
\caption{Regions of the parameter space within reach of various future GW detectors with ${\rm SNR}>10$. Above the solid black curve the supercooling ends with bubble percolation and below by quantum fluctuations. The black dashed curve shows the bound coming from allowed length of the thermal inflation period. Solid brown and red contours indicate, respectively, the exclusions arising from the current LIGO data and PTA observations, while the red dashed contours indicate the regions that fits the recent NANOGrav result with a given confidence level.}
\label{fig:SNRs}
\end{figure}

To systematically study the prospects of GW experiments to probe the above GW signals we perform the standard signal-to-noise (SNR) analysis. The SNR of a given stochastic GW background $\Omega_{\rm GW}(f)$ at a detector whose sensitivity desrcibed by the noise $\Omega_{\rm noise}(f)$ is given by
\be
{\rm SNR} = \sqrt{\mathcal{T} \int \td f \left[\frac{\Omega_{\rm GW}(f)}{\Omega_{\rm noise}(f)}\right]^2} \,,
\ee
where $\mathcal{T}$ denotes the observation period. We asssume $\mathcal{T} = 4\,{\rm yr}$ for all future experiments unless indicated otherwise. In Fig.~\ref{fig:SNRs} we show the ${\rm SNR}>10$ regions for various future GW observatories. Solid lines indicate exclusions we can already put on parts of the parameter space through the current LIGO data and PTA observations while the red region would fit the recent NANOGrav excess with a given significance. Regions with dashed boundaries indicate the sensitivities of upcoming GW experiments. Remarkably, in the near future we will be able to probe almost the entire allowed parameter space.

\section{Conclusions}

We have analysed the end of supercooling in thermal inflation scenarios realised in nearly scale invariant models, focusing on the representative case of a weakly coupled model which allows for an explicit computation of the various possible regimes. We have started with the standard case of bubble nucleation and moving to the case in which the nucleation never proceeds at the rate large enough to end the thermal inflation stage. In the latter case the scalar field fluctuations, building up during the inflationary period, eventually destabilise the false vacuum. After this the exponential expansion still continues for several $e$-folds before the fluctuations become so large that the scalar field finally rolls to the true vacuum. 

The length of the thermal inflation period is bounded by the CMB observations, which could not be correctly reproduced had the appropriate scales exited the horizon during thermal inflation. Specifically, the longer the thermal inflation period lasts the closer to the end of the primordial inflation the CMB scales have to exit the horizon, and we have a hard bound when the second inflation period would last as long as the primordial inflation needs to. We have shown that in large parts of the parameter space of quasi-conformal models the supercooling is escaped by the quantum fluctuations instead of bubble percolation, and that also in such case the thermal inflation period can be short enough to satisfy the CMB bound. 

When the scalar escapes from supercooling by growing quantum fluctuations, the scalar perturbations experience a quite dramatic boost. We have found that the spectrum of curvature perturbations generated during the thermal inflation period can easily grow up to values of order $10^{-4} - 10^{-3}$. Unfortunately this falls slightly short of the amplitudes needed to be able to produce PBHs. It is, however, conceivable that simple extensions of our generic setup might be able to overcome this issue. Nevertheless, such a high amplitude for the power spectrum is enough to generate an observable GW signal.

We have investigated the GW footprint from the quasi-conformal sector for both types of escape routes from the stage of supercooling. In the standard case, where the supercooling ends with bubble percolation, these are produced at the end of the transition as the bubbles collide and field gradients propagate. If nucleation fails and inflationary field fluctuations destabilise the vacuum, GWs are sourced instead by the the scalar fluctuations developing during thermal inflation. We map the GW imprints in both of these cases to the allowed parameter space and find that currently running LIGO and pulsar timing arrays have already excluded a part of region where thermal inflation ends by the bubble percolation. We also find some parameter range where the signal from bubble collisions could explain the excess in NANOGrav data. 

The shapes of the GW spectra in the two cases are distinguishable. The GW spectrum induced by the scalar fluctuations is characterized by a low-frequency peak and an almost flat high-frequency part that can span for several decades, whereas the GW spectrum from bubble collisions is a broken power-law featuring relatively steep tails. We find that the largest GW amplitudes are attained for the parameter values near the separation between the two regimes (with or without bubbles), and the GW signal is stronger in the case that the escape from supercooling was through bubble percolation. Finally, we have shown that, remarkably, the next generation of GW experiments will be able to probe the allowed parameter space almost in its entirety. The possible future detection of a stochastic GW background will then allow us to ascertain whether the Universe underwent supercooling and how it managed to escape from it.

\begin{acknowledgments}
This work was supported by the Spanish MINECO grants FPA2017-88915-P and SEV-2016-0588, the grant 2017-SGR-1069 from the Generalitat de Catalunya, and by the Polish National Science Center grant 2018/31/D/ST2/02048 as well as the Polish National Agency for Academic Exchange within Polish Returns Programme under agreement PPN/PPO/2020/1/00013/U/00001. IFAE is partially funded by the CERCA program of the Generalitat de Catalunya.
\end{acknowledgments}

\bibliography{TI}

\begin{thebibliography}{103}%
\makeatletter
\providecommand \@ifxundefined [1]{%
 \@ifx{#1\undefined}
}%
\providecommand \@ifnum [1]{%
 \ifnum #1\expandafter \@firstoftwo
 \else \expandafter \@secondoftwo
 \fi
}%
\providecommand \@ifx [1]{%
 \ifx #1\expandafter \@firstoftwo
 \else \expandafter \@secondoftwo
 \fi
}%
\providecommand \natexlab [1]{#1}%
\providecommand \enquote  [1]{``#1''}%
\providecommand \bibnamefont  [1]{#1}%
\providecommand \bibfnamefont [1]{#1}%
\providecommand \citenamefont [1]{#1}%
\providecommand \href@noop [0]{\@secondoftwo}%
\providecommand \href [0]{\begingroup \@sanitize@url \@href}%
\providecommand \@href[1]{\@@startlink{#1}\@@href}%
\providecommand \@@href[1]{\endgroup#1\@@endlink}%
\providecommand \@sanitize@url [0]{\catcode `\\12\catcode `\$12\catcode
  `\&12\catcode `\#12\catcode `\^12\catcode `\_12\catcode `\%12\relax}%
\providecommand \@@startlink[1]{}%
\providecommand \@@endlink[0]{}%
\providecommand \url  [0]{\begingroup\@sanitize@url \@url }%
\providecommand \@url [1]{\endgroup\@href {#1}{\urlprefix }}%
\providecommand \urlprefix  [0]{URL }%
\providecommand \Eprint [0]{\href }%
\providecommand \doibase [0]{http://dx.doi.org/}%
\providecommand \selectlanguage [0]{\@gobble}%
\providecommand \bibinfo  [0]{\@secondoftwo}%
\providecommand \bibfield  [0]{\@secondoftwo}%
\providecommand \translation [1]{[#1]}%
\providecommand \BibitemOpen [0]{}%
\providecommand \bibitemStop [0]{}%
\providecommand \bibitemNoStop [0]{.\EOS\space}%
\providecommand \EOS [0]{\spacefactor3000\relax}%
\providecommand \BibitemShut  [1]{\csname bibitem#1\endcsname}%
\let\auto@bib@innerbib\@empty
\bibitem [{\citenamefont {Allahverdi}\ \emph {et~al.}(2020)\citenamefont
  {Allahverdi} \emph {et~al.}}]{Allahverdi:2020bys}%
  \BibitemOpen
  \bibfield  {author} {\bibinfo {author} {\bibfnamefont {R.}~\bibnamefont
  {Allahverdi}} \emph {et~al.},\ }\href {\doibase 10.21105/astro.2006.16182} {\
   (\bibinfo {year} {2020}),\ 10.21105/astro.2006.16182},\ \Eprint
  {http://arxiv.org/abs/2006.16182} {arXiv:2006.16182 [astro-ph.CO]}
  \BibitemShut {NoStop}%
\bibitem [{\citenamefont {Akrami}\ \emph {et~al.}(2020)\citenamefont {Akrami}
  \emph {et~al.}}]{Akrami:2018odb}%
  \BibitemOpen
  \bibfield  {author} {\bibinfo {author} {\bibfnamefont {Y.}~\bibnamefont
  {Akrami}} \emph {et~al.} (\bibinfo {collaboration} {Planck}),\ }\href
  {\doibase 10.1051/0004-6361/201833887} {\bibfield  {journal} {\bibinfo
  {journal} {Astron. Astrophys.}\ }\textbf {\bibinfo {volume} {641}},\ \bibinfo
  {pages} {A10} (\bibinfo {year} {2020})},\ \Eprint
  {http://arxiv.org/abs/1807.06211} {arXiv:1807.06211 [astro-ph.CO]}
  \BibitemShut {NoStop}%
\bibitem [{\citenamefont {Lyth}\ and\ \citenamefont
  {Stewart}(1996)}]{Lyth:1995ka}%
  \BibitemOpen
  \bibfield  {author} {\bibinfo {author} {\bibfnamefont {D.~H.}\ \bibnamefont
  {Lyth}}\ and\ \bibinfo {author} {\bibfnamefont {E.~D.}\ \bibnamefont
  {Stewart}},\ }\href {\doibase 10.1103/PhysRevD.53.1784} {\bibfield  {journal}
  {\bibinfo  {journal} {Phys. Rev. D}\ }\textbf {\bibinfo {volume} {53}},\
  \bibinfo {pages} {1784} (\bibinfo {year} {1996})},\ \Eprint
  {http://arxiv.org/abs/hep-ph/9510204} {arXiv:hep-ph/9510204} \BibitemShut
  {NoStop}%
\bibitem [{\citenamefont {Lyth}\ and\ \citenamefont
  {Stewart}(1995)}]{Lyth:1995hj}%
  \BibitemOpen
  \bibfield  {author} {\bibinfo {author} {\bibfnamefont {D.~H.}\ \bibnamefont
  {Lyth}}\ and\ \bibinfo {author} {\bibfnamefont {E.~D.}\ \bibnamefont
  {Stewart}},\ }\href {\doibase 10.1103/PhysRevLett.75.201} {\bibfield
  {journal} {\bibinfo  {journal} {Phys. Rev. Lett.}\ }\textbf {\bibinfo
  {volume} {75}},\ \bibinfo {pages} {201} (\bibinfo {year} {1995})},\ \Eprint
  {http://arxiv.org/abs/hep-ph/9502417} {arXiv:hep-ph/9502417} \BibitemShut
  {NoStop}%
\bibitem [{\citenamefont {Witten}(1981)}]{Witten:1980ez}%
  \BibitemOpen
  \bibfield  {author} {\bibinfo {author} {\bibfnamefont {E.}~\bibnamefont
  {Witten}},\ }\href {\doibase 10.1016/0550-3213(81)90182-6} {\bibfield
  {journal} {\bibinfo  {journal} {Nucl. Phys. B}\ }\textbf {\bibinfo {volume}
  {177}},\ \bibinfo {pages} {477} (\bibinfo {year} {1981})}\BibitemShut
  {NoStop}%
\bibitem [{\citenamefont {Quigg}\ and\ \citenamefont
  {Shrock}(2009)}]{Quigg:2009xr}%
  \BibitemOpen
  \bibfield  {author} {\bibinfo {author} {\bibfnamefont {C.}~\bibnamefont
  {Quigg}}\ and\ \bibinfo {author} {\bibfnamefont {R.}~\bibnamefont {Shrock}},\
  }\href {\doibase 10.1103/PhysRevD.79.096002} {\bibfield  {journal} {\bibinfo
  {journal} {Phys. Rev. D}\ }\textbf {\bibinfo {volume} {79}},\ \bibinfo
  {pages} {096002} (\bibinfo {year} {2009})},\ \Eprint
  {http://arxiv.org/abs/0901.3958} {arXiv:0901.3958 [hep-ph]} \BibitemShut
  {NoStop}%
\bibitem [{\citenamefont {Iso}\ \emph {et~al.}(2017)\citenamefont {Iso},
  \citenamefont {Serpico},\ and\ \citenamefont {Shimada}}]{Iso:2017uuu}%
  \BibitemOpen
  \bibfield  {author} {\bibinfo {author} {\bibfnamefont {S.}~\bibnamefont
  {Iso}}, \bibinfo {author} {\bibfnamefont {P.~D.}\ \bibnamefont {Serpico}}, \
  and\ \bibinfo {author} {\bibfnamefont {K.}~\bibnamefont {Shimada}},\ }\href
  {\doibase 10.1103/PhysRevLett.119.141301} {\bibfield  {journal} {\bibinfo
  {journal} {Phys. Rev. Lett.}\ }\textbf {\bibinfo {volume} {119}},\ \bibinfo
  {pages} {141301} (\bibinfo {year} {2017})},\ \Eprint
  {http://arxiv.org/abs/1704.04955} {arXiv:1704.04955 [hep-ph]} \BibitemShut
  {NoStop}%
\bibitem [{\citenamefont {Hambye}\ \emph {et~al.}(2018)\citenamefont {Hambye},
  \citenamefont {Strumia},\ and\ \citenamefont {Teresi}}]{Hambye:2018qjv}%
  \BibitemOpen
  \bibfield  {author} {\bibinfo {author} {\bibfnamefont {T.}~\bibnamefont
  {Hambye}}, \bibinfo {author} {\bibfnamefont {A.}~\bibnamefont {Strumia}}, \
  and\ \bibinfo {author} {\bibfnamefont {D.}~\bibnamefont {Teresi}},\ }\href
  {\doibase 10.1007/JHEP08(2018)188} {\bibfield  {journal} {\bibinfo  {journal}
  {JHEP}\ }\textbf {\bibinfo {volume} {08}},\ \bibinfo {pages} {188} (\bibinfo
  {year} {2018})},\ \Eprint {http://arxiv.org/abs/1805.01473} {arXiv:1805.01473
  [hep-ph]} \BibitemShut {NoStop}%
\bibitem [{\citenamefont {von Harling}\ and\ \citenamefont
  {Servant}(2018)}]{vonHarling:2017yew}%
  \BibitemOpen
  \bibfield  {author} {\bibinfo {author} {\bibfnamefont {B.}~\bibnamefont {von
  Harling}}\ and\ \bibinfo {author} {\bibfnamefont {G.}~\bibnamefont
  {Servant}},\ }\href {\doibase 10.1007/JHEP01(2018)159} {\bibfield  {journal}
  {\bibinfo  {journal} {JHEP}\ }\textbf {\bibinfo {volume} {01}},\ \bibinfo
  {pages} {159} (\bibinfo {year} {2018})},\ \Eprint
  {http://arxiv.org/abs/1711.11554} {arXiv:1711.11554 [hep-ph]} \BibitemShut
  {NoStop}%
\bibitem [{\citenamefont {Marzo}\ \emph {et~al.}(2019)\citenamefont {Marzo},
  \citenamefont {Marzola},\ and\ \citenamefont {Vaskonen}}]{Marzo:2018nov}%
  \BibitemOpen
  \bibfield  {author} {\bibinfo {author} {\bibfnamefont {C.}~\bibnamefont
  {Marzo}}, \bibinfo {author} {\bibfnamefont {L.}~\bibnamefont {Marzola}}, \
  and\ \bibinfo {author} {\bibfnamefont {V.}~\bibnamefont {Vaskonen}},\ }\href
  {\doibase 10.1140/epjc/s10052-019-7076-x} {\bibfield  {journal} {\bibinfo
  {journal} {Eur. Phys. J. C}\ }\textbf {\bibinfo {volume} {79}},\ \bibinfo
  {pages} {601} (\bibinfo {year} {2019})},\ \Eprint
  {http://arxiv.org/abs/1811.11169} {arXiv:1811.11169 [hep-ph]} \BibitemShut
  {NoStop}%
\bibitem [{\citenamefont {Baratella}\ \emph {et~al.}(2019)\citenamefont
  {Baratella}, \citenamefont {Pomarol},\ and\ \citenamefont
  {Rompineve}}]{Baratella:2018pxi}%
  \BibitemOpen
  \bibfield  {author} {\bibinfo {author} {\bibfnamefont {P.}~\bibnamefont
  {Baratella}}, \bibinfo {author} {\bibfnamefont {A.}~\bibnamefont {Pomarol}},
  \ and\ \bibinfo {author} {\bibfnamefont {F.}~\bibnamefont {Rompineve}},\
  }\href {\doibase 10.1007/JHEP03(2019)100} {\bibfield  {journal} {\bibinfo
  {journal} {JHEP}\ }\textbf {\bibinfo {volume} {03}},\ \bibinfo {pages} {100}
  (\bibinfo {year} {2019})},\ \Eprint {http://arxiv.org/abs/1812.06996}
  {arXiv:1812.06996 [hep-ph]} \BibitemShut {NoStop}%
\bibitem [{\citenamefont {Vilenkin}\ and\ \citenamefont
  {Ford}(1982)}]{Vilenkin:1982wt}%
  \BibitemOpen
  \bibfield  {author} {\bibinfo {author} {\bibfnamefont {A.}~\bibnamefont
  {Vilenkin}}\ and\ \bibinfo {author} {\bibfnamefont {L.~H.}\ \bibnamefont
  {Ford}},\ }\href {\doibase 10.1103/PhysRevD.26.1231} {\bibfield  {journal}
  {\bibinfo  {journal} {Phys. Rev. D}\ }\textbf {\bibinfo {volume} {26}},\
  \bibinfo {pages} {1231} (\bibinfo {year} {1982})}\BibitemShut {NoStop}%
\bibitem [{\citenamefont {Vilenkin}(1983{\natexlab{a}})}]{Vilenkin:1982sg}%
  \BibitemOpen
  \bibfield  {author} {\bibinfo {author} {\bibfnamefont {A.}~\bibnamefont
  {Vilenkin}},\ }\href {\doibase 10.1016/0550-3213(83)90207-9} {\bibfield
  {journal} {\bibinfo  {journal} {Nucl. Phys. B}\ }\textbf {\bibinfo {volume}
  {226}},\ \bibinfo {pages} {504} (\bibinfo {year}
  {1983}{\natexlab{a}})}\BibitemShut {NoStop}%
\bibitem [{\citenamefont {Vilenkin}(1983{\natexlab{b}})}]{Vilenkin:1983xp}%
  \BibitemOpen
  \bibfield  {author} {\bibinfo {author} {\bibfnamefont {A.}~\bibnamefont
  {Vilenkin}},\ }\href {\doibase 10.1016/0550-3213(83)90208-0} {\bibfield
  {journal} {\bibinfo  {journal} {Nucl. Phys. B}\ }\textbf {\bibinfo {volume}
  {226}},\ \bibinfo {pages} {527} (\bibinfo {year}
  {1983}{\natexlab{b}})}\BibitemShut {NoStop}%
\bibitem [{\citenamefont {Creminelli}\ \emph {et~al.}(2002)\citenamefont
  {Creminelli}, \citenamefont {Nicolis},\ and\ \citenamefont
  {Rattazzi}}]{Creminelli:2001th}%
  \BibitemOpen
  \bibfield  {author} {\bibinfo {author} {\bibfnamefont {P.}~\bibnamefont
  {Creminelli}}, \bibinfo {author} {\bibfnamefont {A.}~\bibnamefont {Nicolis}},
  \ and\ \bibinfo {author} {\bibfnamefont {R.}~\bibnamefont {Rattazzi}},\
  }\href {\doibase 10.1088/1126-6708/2002/03/051} {\bibfield  {journal}
  {\bibinfo  {journal} {JHEP}\ }\textbf {\bibinfo {volume} {03}},\ \bibinfo
  {pages} {051} (\bibinfo {year} {2002})},\ \Eprint
  {http://arxiv.org/abs/hep-th/0107141} {arXiv:hep-th/0107141} \BibitemShut
  {NoStop}%
\bibitem [{\citenamefont {Randall}\ and\ \citenamefont
  {Servant}(2007)}]{Randall:2006py}%
  \BibitemOpen
  \bibfield  {author} {\bibinfo {author} {\bibfnamefont {L.}~\bibnamefont
  {Randall}}\ and\ \bibinfo {author} {\bibfnamefont {G.}~\bibnamefont
  {Servant}},\ }\href {\doibase 10.1088/1126-6708/2007/05/054} {\bibfield
  {journal} {\bibinfo  {journal} {JHEP}\ }\textbf {\bibinfo {volume} {05}},\
  \bibinfo {pages} {054} (\bibinfo {year} {2007})},\ \Eprint
  {http://arxiv.org/abs/hep-ph/0607158} {arXiv:hep-ph/0607158} \BibitemShut
  {NoStop}%
\bibitem [{\citenamefont {Konstandin}\ and\ \citenamefont
  {Servant}(2011{\natexlab{a}})}]{Konstandin:2011ds}%
  \BibitemOpen
  \bibfield  {author} {\bibinfo {author} {\bibfnamefont {T.}~\bibnamefont
  {Konstandin}}\ and\ \bibinfo {author} {\bibfnamefont {G.}~\bibnamefont
  {Servant}},\ }\href {\doibase 10.1088/1475-7516/2011/07/024} {\bibfield
  {journal} {\bibinfo  {journal} {JCAP}\ }\textbf {\bibinfo {volume} {07}},\
  \bibinfo {pages} {024} (\bibinfo {year} {2011}{\natexlab{a}})},\ \Eprint
  {http://arxiv.org/abs/1104.4793} {arXiv:1104.4793 [hep-ph]} \BibitemShut
  {NoStop}%
\bibitem [{\citenamefont {Nardini}\ \emph {et~al.}(2007)\citenamefont
  {Nardini}, \citenamefont {Quiros},\ and\ \citenamefont
  {Wulzer}}]{Nardini:2007me}%
  \BibitemOpen
  \bibfield  {author} {\bibinfo {author} {\bibfnamefont {G.}~\bibnamefont
  {Nardini}}, \bibinfo {author} {\bibfnamefont {M.}~\bibnamefont {Quiros}}, \
  and\ \bibinfo {author} {\bibfnamefont {A.}~\bibnamefont {Wulzer}},\ }\href
  {\doibase 10.1088/1126-6708/2007/09/077} {\bibfield  {journal} {\bibinfo
  {journal} {JHEP}\ }\textbf {\bibinfo {volume} {09}},\ \bibinfo {pages} {077}
  (\bibinfo {year} {2007})},\ \Eprint {http://arxiv.org/abs/0706.3388}
  {arXiv:0706.3388 [hep-ph]} \BibitemShut {NoStop}%
\bibitem [{\citenamefont {Konstandin}\ and\ \citenamefont
  {Servant}(2011{\natexlab{b}})}]{Konstandin:2011dr}%
  \BibitemOpen
  \bibfield  {author} {\bibinfo {author} {\bibfnamefont {T.}~\bibnamefont
  {Konstandin}}\ and\ \bibinfo {author} {\bibfnamefont {G.}~\bibnamefont
  {Servant}},\ }\href {\doibase 10.1088/1475-7516/2011/12/009} {\bibfield
  {journal} {\bibinfo  {journal} {JCAP}\ }\textbf {\bibinfo {volume} {12}},\
  \bibinfo {pages} {009} (\bibinfo {year} {2011}{\natexlab{b}})},\ \Eprint
  {http://arxiv.org/abs/1104.4791} {arXiv:1104.4791 [hep-ph]} \BibitemShut
  {NoStop}%
\bibitem [{\citenamefont {Bunk}\ \emph {et~al.}(2018)\citenamefont {Bunk},
  \citenamefont {Hubisz},\ and\ \citenamefont {Jain}}]{Bunk:2017fic}%
  \BibitemOpen
  \bibfield  {author} {\bibinfo {author} {\bibfnamefont {D.}~\bibnamefont
  {Bunk}}, \bibinfo {author} {\bibfnamefont {J.}~\bibnamefont {Hubisz}}, \ and\
  \bibinfo {author} {\bibfnamefont {B.}~\bibnamefont {Jain}},\ }\href {\doibase
  10.1140/epjc/s10052-018-5529-2} {\bibfield  {journal} {\bibinfo  {journal}
  {Eur. Phys. J. C}\ }\textbf {\bibinfo {volume} {78}},\ \bibinfo {pages} {78}
  (\bibinfo {year} {2018})},\ \Eprint {http://arxiv.org/abs/1705.00001}
  {arXiv:1705.00001 [hep-ph]} \BibitemShut {NoStop}%
\bibitem [{\citenamefont {Dillon}\ \emph {et~al.}(2018)\citenamefont {Dillon},
  \citenamefont {El-Menoufi}, \citenamefont {Huber},\ and\ \citenamefont
  {Manuel}}]{Dillon:2017ctw}%
  \BibitemOpen
  \bibfield  {author} {\bibinfo {author} {\bibfnamefont {B.~M.}\ \bibnamefont
  {Dillon}}, \bibinfo {author} {\bibfnamefont {B.~K.}\ \bibnamefont
  {El-Menoufi}}, \bibinfo {author} {\bibfnamefont {S.~J.}\ \bibnamefont
  {Huber}}, \ and\ \bibinfo {author} {\bibfnamefont {J.~P.}\ \bibnamefont
  {Manuel}},\ }\href {\doibase 10.1103/PhysRevD.98.086005} {\bibfield
  {journal} {\bibinfo  {journal} {Phys. Rev. D}\ }\textbf {\bibinfo {volume}
  {98}},\ \bibinfo {pages} {086005} (\bibinfo {year} {2018})},\ \Eprint
  {http://arxiv.org/abs/1708.02953} {arXiv:1708.02953 [hep-th]} \BibitemShut
  {NoStop}%
\bibitem [{\citenamefont {Meg\'\i{}as}\ \emph {et~al.}(2018)\citenamefont
  {Meg\'\i{}as}, \citenamefont {Nardini},\ and\ \citenamefont
  {Quir\'os}}]{Megias:2018sxv}%
  \BibitemOpen
  \bibfield  {author} {\bibinfo {author} {\bibfnamefont {E.}~\bibnamefont
  {Meg\'\i{}as}}, \bibinfo {author} {\bibfnamefont {G.}~\bibnamefont
  {Nardini}}, \ and\ \bibinfo {author} {\bibfnamefont {M.}~\bibnamefont
  {Quir\'os}},\ }\href {\doibase 10.1007/JHEP09(2018)095} {\bibfield  {journal}
  {\bibinfo  {journal} {JHEP}\ }\textbf {\bibinfo {volume} {09}},\ \bibinfo
  {pages} {095} (\bibinfo {year} {2018})},\ \Eprint
  {http://arxiv.org/abs/1806.04877} {arXiv:1806.04877 [hep-ph]} \BibitemShut
  {NoStop}%
\bibitem [{\citenamefont {Bruggisser}\ \emph {et~al.}(2018)\citenamefont
  {Bruggisser}, \citenamefont {Von~Harling}, \citenamefont {Matsedonskyi},\
  and\ \citenamefont {Servant}}]{Bruggisser:2018mrt}%
  \BibitemOpen
  \bibfield  {author} {\bibinfo {author} {\bibfnamefont {S.}~\bibnamefont
  {Bruggisser}}, \bibinfo {author} {\bibfnamefont {B.}~\bibnamefont
  {Von~Harling}}, \bibinfo {author} {\bibfnamefont {O.}~\bibnamefont
  {Matsedonskyi}}, \ and\ \bibinfo {author} {\bibfnamefont {G.}~\bibnamefont
  {Servant}},\ }\href {\doibase 10.1007/JHEP12(2018)099} {\bibfield  {journal}
  {\bibinfo  {journal} {JHEP}\ }\textbf {\bibinfo {volume} {12}},\ \bibinfo
  {pages} {099} (\bibinfo {year} {2018})},\ \Eprint
  {http://arxiv.org/abs/1804.07314} {arXiv:1804.07314 [hep-ph]} \BibitemShut
  {NoStop}%
\bibitem [{\citenamefont {Aoki}\ and\ \citenamefont
  {Kubo}(2020)}]{Aoki:2019mlt}%
  \BibitemOpen
  \bibfield  {author} {\bibinfo {author} {\bibfnamefont {M.}~\bibnamefont
  {Aoki}}\ and\ \bibinfo {author} {\bibfnamefont {J.}~\bibnamefont {Kubo}},\
  }\href {\doibase 10.1088/1475-7516/2020/04/001} {\bibfield  {journal}
  {\bibinfo  {journal} {JCAP}\ }\textbf {\bibinfo {volume} {04}},\ \bibinfo
  {pages} {001} (\bibinfo {year} {2020})},\ \Eprint
  {http://arxiv.org/abs/1910.05025} {arXiv:1910.05025 [hep-ph]} \BibitemShut
  {NoStop}%
\bibitem [{\citenamefont {Von~Harling}\ \emph {et~al.}(2020)\citenamefont
  {Von~Harling}, \citenamefont {Pomarol}, \citenamefont {Pujol\`as},\ and\
  \citenamefont {Rompineve}}]{vonHarling:2019gme}%
  \BibitemOpen
  \bibfield  {author} {\bibinfo {author} {\bibfnamefont {B.}~\bibnamefont
  {Von~Harling}}, \bibinfo {author} {\bibfnamefont {A.}~\bibnamefont
  {Pomarol}}, \bibinfo {author} {\bibfnamefont {O.}~\bibnamefont {Pujol\`as}},
  \ and\ \bibinfo {author} {\bibfnamefont {F.}~\bibnamefont {Rompineve}},\
  }\href {\doibase 10.1007/JHEP04(2020)195} {\bibfield  {journal} {\bibinfo
  {journal} {JHEP}\ }\textbf {\bibinfo {volume} {04}},\ \bibinfo {pages} {195}
  (\bibinfo {year} {2020})},\ \Eprint {http://arxiv.org/abs/1912.07587}
  {arXiv:1912.07587 [hep-ph]} \BibitemShut {NoStop}%
\bibitem [{\citenamefont {Delle~Rose}\ \emph {et~al.}(2020)\citenamefont
  {Delle~Rose}, \citenamefont {Panico}, \citenamefont {Redi},\ and\
  \citenamefont {Tesi}}]{DelleRose:2019pgi}%
  \BibitemOpen
  \bibfield  {author} {\bibinfo {author} {\bibfnamefont {L.}~\bibnamefont
  {Delle~Rose}}, \bibinfo {author} {\bibfnamefont {G.}~\bibnamefont {Panico}},
  \bibinfo {author} {\bibfnamefont {M.}~\bibnamefont {Redi}}, \ and\ \bibinfo
  {author} {\bibfnamefont {A.}~\bibnamefont {Tesi}},\ }\href {\doibase
  10.1007/JHEP04(2020)025} {\bibfield  {journal} {\bibinfo  {journal} {JHEP}\
  }\textbf {\bibinfo {volume} {04}},\ \bibinfo {pages} {025} (\bibinfo {year}
  {2020})},\ \Eprint {http://arxiv.org/abs/1912.06139} {arXiv:1912.06139
  [hep-ph]} \BibitemShut {NoStop}%
\bibitem [{\citenamefont {Azatov}\ and\ \citenamefont
  {Vanvlasselaer}(2020)}]{Azatov:2020nbe}%
  \BibitemOpen
  \bibfield  {author} {\bibinfo {author} {\bibfnamefont {A.}~\bibnamefont
  {Azatov}}\ and\ \bibinfo {author} {\bibfnamefont {M.}~\bibnamefont
  {Vanvlasselaer}},\ }\href {\doibase 10.1007/JHEP09(2020)085} {\bibfield
  {journal} {\bibinfo  {journal} {JHEP}\ }\textbf {\bibinfo {volume} {09}},\
  \bibinfo {pages} {085} (\bibinfo {year} {2020})},\ \Eprint
  {http://arxiv.org/abs/2003.10265} {arXiv:2003.10265 [hep-ph]} \BibitemShut
  {NoStop}%
\bibitem [{\citenamefont {Coleman}\ and\ \citenamefont
  {Weinberg}(1973)}]{Coleman:1973jx}%
  \BibitemOpen
  \bibfield  {author} {\bibinfo {author} {\bibfnamefont {S.~R.}\ \bibnamefont
  {Coleman}}\ and\ \bibinfo {author} {\bibfnamefont {E.~J.}\ \bibnamefont
  {Weinberg}},\ }\href {\doibase 10.1103/PhysRevD.7.1888} {\bibfield  {journal}
  {\bibinfo  {journal} {Phys. Rev. D}\ }\textbf {\bibinfo {volume} {7}},\
  \bibinfo {pages} {1888} (\bibinfo {year} {1973})}\BibitemShut {NoStop}%
\bibitem [{\citenamefont {Jinno}\ and\ \citenamefont
  {Takimoto}(2017)}]{Jinno:2016knw}%
  \BibitemOpen
  \bibfield  {author} {\bibinfo {author} {\bibfnamefont {R.}~\bibnamefont
  {Jinno}}\ and\ \bibinfo {author} {\bibfnamefont {M.}~\bibnamefont
  {Takimoto}},\ }\href {\doibase 10.1103/PhysRevD.95.015020} {\bibfield
  {journal} {\bibinfo  {journal} {Phys. Rev. D}\ }\textbf {\bibinfo {volume}
  {95}},\ \bibinfo {pages} {015020} (\bibinfo {year} {2017})},\ \Eprint
  {http://arxiv.org/abs/1604.05035} {arXiv:1604.05035 [hep-ph]} \BibitemShut
  {NoStop}%
\bibitem [{\citenamefont {Jaeckel}\ \emph {et~al.}(2016)\citenamefont
  {Jaeckel}, \citenamefont {Khoze},\ and\ \citenamefont
  {Spannowsky}}]{Jaeckel:2016jlh}%
  \BibitemOpen
  \bibfield  {author} {\bibinfo {author} {\bibfnamefont {J.}~\bibnamefont
  {Jaeckel}}, \bibinfo {author} {\bibfnamefont {V.~V.}\ \bibnamefont {Khoze}},
  \ and\ \bibinfo {author} {\bibfnamefont {M.}~\bibnamefont {Spannowsky}},\
  }\href {\doibase 10.1103/PhysRevD.94.103519} {\bibfield  {journal} {\bibinfo
  {journal} {Phys. Rev. D}\ }\textbf {\bibinfo {volume} {94}},\ \bibinfo
  {pages} {103519} (\bibinfo {year} {2016})},\ \Eprint
  {http://arxiv.org/abs/1602.03901} {arXiv:1602.03901 [hep-ph]} \BibitemShut
  {NoStop}%
\bibitem [{\citenamefont {Marzola}\ \emph {et~al.}(2017)\citenamefont
  {Marzola}, \citenamefont {Racioppi},\ and\ \citenamefont
  {Vaskonen}}]{Marzola:2017jzl}%
  \BibitemOpen
  \bibfield  {author} {\bibinfo {author} {\bibfnamefont {L.}~\bibnamefont
  {Marzola}}, \bibinfo {author} {\bibfnamefont {A.}~\bibnamefont {Racioppi}}, \
  and\ \bibinfo {author} {\bibfnamefont {V.}~\bibnamefont {Vaskonen}},\ }\href
  {\doibase 10.1140/epjc/s10052-017-4996-1} {\bibfield  {journal} {\bibinfo
  {journal} {Eur. Phys. J. C}\ }\textbf {\bibinfo {volume} {77}},\ \bibinfo
  {pages} {484} (\bibinfo {year} {2017})},\ \Eprint
  {http://arxiv.org/abs/1704.01034} {arXiv:1704.01034 [hep-ph]} \BibitemShut
  {NoStop}%
\bibitem [{\citenamefont {Prokopec}\ \emph {et~al.}(2019)\citenamefont
  {Prokopec}, \citenamefont {Rezacek},\ and\ \citenamefont
  {\'Swie\.zewska}}]{Prokopec:2018tnq}%
  \BibitemOpen
  \bibfield  {author} {\bibinfo {author} {\bibfnamefont {T.}~\bibnamefont
  {Prokopec}}, \bibinfo {author} {\bibfnamefont {J.}~\bibnamefont {Rezacek}}, \
  and\ \bibinfo {author} {\bibfnamefont {B.}~\bibnamefont {\'Swie\.zewska}},\
  }\href {\doibase 10.1088/1475-7516/2019/02/009} {\bibfield  {journal}
  {\bibinfo  {journal} {JCAP}\ }\textbf {\bibinfo {volume} {02}},\ \bibinfo
  {pages} {009} (\bibinfo {year} {2019})},\ \Eprint
  {http://arxiv.org/abs/1809.11129} {arXiv:1809.11129 [hep-ph]} \BibitemShut
  {NoStop}%
\bibitem [{\citenamefont {Baldes}\ and\ \citenamefont
  {Garcia-Cely}(2019)}]{Baldes:2018emh}%
  \BibitemOpen
  \bibfield  {author} {\bibinfo {author} {\bibfnamefont {I.}~\bibnamefont
  {Baldes}}\ and\ \bibinfo {author} {\bibfnamefont {C.}~\bibnamefont
  {Garcia-Cely}},\ }\href {\doibase 10.1007/JHEP05(2019)190} {\bibfield
  {journal} {\bibinfo  {journal} {JHEP}\ }\textbf {\bibinfo {volume} {05}},\
  \bibinfo {pages} {190} (\bibinfo {year} {2019})},\ \Eprint
  {http://arxiv.org/abs/1809.01198} {arXiv:1809.01198 [hep-ph]} \BibitemShut
  {NoStop}%
\bibitem [{\citenamefont {Fujikura}\ \emph {et~al.}(2020)\citenamefont
  {Fujikura}, \citenamefont {Nakai},\ and\ \citenamefont
  {Yamada}}]{Fujikura:2019oyi}%
  \BibitemOpen
  \bibfield  {author} {\bibinfo {author} {\bibfnamefont {K.}~\bibnamefont
  {Fujikura}}, \bibinfo {author} {\bibfnamefont {Y.}~\bibnamefont {Nakai}}, \
  and\ \bibinfo {author} {\bibfnamefont {M.}~\bibnamefont {Yamada}},\ }\href
  {\doibase 10.1007/JHEP02(2020)111} {\bibfield  {journal} {\bibinfo  {journal}
  {JHEP}\ }\textbf {\bibinfo {volume} {02}},\ \bibinfo {pages} {111} (\bibinfo
  {year} {2020})},\ \Eprint {http://arxiv.org/abs/1910.07546} {arXiv:1910.07546
  [hep-ph]} \BibitemShut {NoStop}%
\bibitem [{\citenamefont {Wang}\ \emph {et~al.}(2020)\citenamefont {Wang},
  \citenamefont {Huang},\ and\ \citenamefont {Zhang}}]{Wang:2020jrd}%
  \BibitemOpen
  \bibfield  {author} {\bibinfo {author} {\bibfnamefont {X.}~\bibnamefont
  {Wang}}, \bibinfo {author} {\bibfnamefont {F.~P.}\ \bibnamefont {Huang}}, \
  and\ \bibinfo {author} {\bibfnamefont {X.}~\bibnamefont {Zhang}},\ }\href
  {\doibase 10.1088/1475-7516/2020/05/045} {\bibfield  {journal} {\bibinfo
  {journal} {JCAP}\ }\textbf {\bibinfo {volume} {05}},\ \bibinfo {pages} {045}
  (\bibinfo {year} {2020})},\ \Eprint {http://arxiv.org/abs/2003.08892}
  {arXiv:2003.08892 [hep-ph]} \BibitemShut {NoStop}%
\bibitem [{\citenamefont {Ellis}\ \emph {et~al.}(2020)\citenamefont {Ellis},
  \citenamefont {Lewicki},\ and\ \citenamefont {Vaskonen}}]{Ellis:2020nnr}%
  \BibitemOpen
  \bibfield  {author} {\bibinfo {author} {\bibfnamefont {J.}~\bibnamefont
  {Ellis}}, \bibinfo {author} {\bibfnamefont {M.}~\bibnamefont {Lewicki}}, \
  and\ \bibinfo {author} {\bibfnamefont {V.}~\bibnamefont {Vaskonen}},\ }\href
  {\doibase 10.1088/1475-7516/2020/11/020} {\bibfield  {journal} {\bibinfo
  {journal} {JCAP}\ }\textbf {\bibinfo {volume} {11}},\ \bibinfo {pages} {020}
  (\bibinfo {year} {2020})},\ \Eprint {http://arxiv.org/abs/2007.15586}
  {arXiv:2007.15586 [astro-ph.CO]} \BibitemShut {NoStop}%
\bibitem [{\citenamefont {Aghanim}\ \emph {et~al.}(2020)\citenamefont {Aghanim}
  \emph {et~al.}}]{Aghanim:2018eyx}%
  \BibitemOpen
  \bibfield  {author} {\bibinfo {author} {\bibfnamefont {N.}~\bibnamefont
  {Aghanim}} \emph {et~al.} (\bibinfo {collaboration} {Planck}),\ }\href
  {\doibase 10.1051/0004-6361/201833910} {\bibfield  {journal} {\bibinfo
  {journal} {Astron. Astrophys.}\ }\textbf {\bibinfo {volume} {641}},\ \bibinfo
  {pages} {A6} (\bibinfo {year} {2020})},\ \Eprint
  {http://arxiv.org/abs/1807.06209} {arXiv:1807.06209 [astro-ph.CO]}
  \BibitemShut {NoStop}%
\bibitem [{\citenamefont {Linde}(1983)}]{Linde:1981zj}%
  \BibitemOpen
  \bibfield  {author} {\bibinfo {author} {\bibfnamefont {A.~D.}\ \bibnamefont
  {Linde}},\ }\href {\doibase 10.1016/0550-3213(83)90072-X} {\bibfield
  {journal} {\bibinfo  {journal} {Nucl. Phys. B}\ }\textbf {\bibinfo {volume}
  {216}},\ \bibinfo {pages} {421} (\bibinfo {year} {1983})},\ \bibinfo {note}
  {[Erratum: Nucl.Phys.B 223, 544 (1983)]}\BibitemShut {NoStop}%
\bibitem [{\citenamefont {Guth}\ and\ \citenamefont {Tye}(1980)}]{Guth:1979bh}%
  \BibitemOpen
  \bibfield  {author} {\bibinfo {author} {\bibfnamefont {A.~H.}\ \bibnamefont
  {Guth}}\ and\ \bibinfo {author} {\bibfnamefont {S.~H.~H.}\ \bibnamefont
  {Tye}},\ }\href {\doibase 10.1103/PhysRevLett.44.631} {\bibfield  {journal}
  {\bibinfo  {journal} {Phys. Rev. Lett.}\ }\textbf {\bibinfo {volume} {44}},\
  \bibinfo {pages} {631} (\bibinfo {year} {1980})},\ \bibinfo {note} {[Erratum:
  Phys.Rev.Lett. 44, 963 (1980)]}\BibitemShut {NoStop}%
\bibitem [{\citenamefont {Guth}\ and\ \citenamefont
  {Weinberg}(1981)}]{Guth:1981uk}%
  \BibitemOpen
  \bibfield  {author} {\bibinfo {author} {\bibfnamefont {A.~H.}\ \bibnamefont
  {Guth}}\ and\ \bibinfo {author} {\bibfnamefont {E.~J.}\ \bibnamefont
  {Weinberg}},\ }\href {\doibase 10.1103/PhysRevD.23.876} {\bibfield  {journal}
  {\bibinfo  {journal} {Phys. Rev. D}\ }\textbf {\bibinfo {volume} {23}},\
  \bibinfo {pages} {876} (\bibinfo {year} {1981})}\BibitemShut {NoStop}%
\bibitem [{\citenamefont {Ellis}\ \emph
  {et~al.}(2019{\natexlab{a}})\citenamefont {Ellis}, \citenamefont {Lewicki},\
  and\ \citenamefont {No}}]{Ellis:2018mja}%
  \BibitemOpen
  \bibfield  {author} {\bibinfo {author} {\bibfnamefont {J.}~\bibnamefont
  {Ellis}}, \bibinfo {author} {\bibfnamefont {M.}~\bibnamefont {Lewicki}}, \
  and\ \bibinfo {author} {\bibfnamefont {J.~M.}\ \bibnamefont {No}},\ }\href
  {\doibase 10.1088/1475-7516/2019/04/003} {\bibfield  {journal} {\bibinfo
  {journal} {JCAP}\ }\textbf {\bibinfo {volume} {04}},\ \bibinfo {pages} {003}
  (\bibinfo {year} {2019}{\natexlab{a}})},\ \Eprint
  {http://arxiv.org/abs/1809.08242} {arXiv:1809.08242 [hep-ph]} \BibitemShut
  {NoStop}%
\bibitem [{\citenamefont {Coleman}\ and\ \citenamefont
  {De~Luccia}(1980)}]{Coleman:1980aw}%
  \BibitemOpen
  \bibfield  {author} {\bibinfo {author} {\bibfnamefont {S.~R.}\ \bibnamefont
  {Coleman}}\ and\ \bibinfo {author} {\bibfnamefont {F.}~\bibnamefont
  {De~Luccia}},\ }\href {\doibase 10.1103/PhysRevD.21.3305} {\bibfield
  {journal} {\bibinfo  {journal} {Phys. Rev. D}\ }\textbf {\bibinfo {volume}
  {21}},\ \bibinfo {pages} {3305} (\bibinfo {year} {1980})}\BibitemShut
  {NoStop}%
\bibitem [{\citenamefont {Czerwi\'nska}\ \emph {et~al.}(2016)\citenamefont
  {Czerwi\'nska}, \citenamefont {Lalak}, \citenamefont {Lewicki},\ and\
  \citenamefont {Olszewski}}]{Czerwinska:2016fky}%
  \BibitemOpen
  \bibfield  {author} {\bibinfo {author} {\bibfnamefont {O.}~\bibnamefont
  {Czerwi\'nska}}, \bibinfo {author} {\bibfnamefont {Z.}~\bibnamefont {Lalak}},
  \bibinfo {author} {\bibfnamefont {M.}~\bibnamefont {Lewicki}}, \ and\
  \bibinfo {author} {\bibfnamefont {P.}~\bibnamefont {Olszewski}},\ }\href
  {\doibase 10.1007/JHEP10(2016)004} {\bibfield  {journal} {\bibinfo  {journal}
  {JHEP}\ }\textbf {\bibinfo {volume} {10}},\ \bibinfo {pages} {004} (\bibinfo
  {year} {2016})},\ \Eprint {http://arxiv.org/abs/1606.07808} {arXiv:1606.07808
  [hep-ph]} \BibitemShut {NoStop}%
\bibitem [{\citenamefont {Bunch}\ and\ \citenamefont
  {Davies}(1978)}]{Bunch:1978yq}%
  \BibitemOpen
  \bibfield  {author} {\bibinfo {author} {\bibfnamefont {T.~S.}\ \bibnamefont
  {Bunch}}\ and\ \bibinfo {author} {\bibfnamefont {P.~C.~W.}\ \bibnamefont
  {Davies}},\ }\href {\doibase 10.1098/rspa.1978.0060} {\bibfield  {journal}
  {\bibinfo  {journal} {Proc. Roy. Soc. Lond. A}\ }\textbf {\bibinfo {volume}
  {360}},\ \bibinfo {pages} {117} (\bibinfo {year} {1978})}\BibitemShut
  {NoStop}%
\bibitem [{\citenamefont {Linde}(1982)}]{Linde:1982uu}%
  \BibitemOpen
  \bibfield  {author} {\bibinfo {author} {\bibfnamefont {A.~D.}\ \bibnamefont
  {Linde}},\ }\href {\doibase 10.1016/0370-2693(82)90293-3} {\bibfield
  {journal} {\bibinfo  {journal} {Phys. Lett. B}\ }\textbf {\bibinfo {volume}
  {116}},\ \bibinfo {pages} {335} (\bibinfo {year} {1982})}\BibitemShut
  {NoStop}%
\bibitem [{\citenamefont {Starobinsky}(1982)}]{Starobinsky:1982ee}%
  \BibitemOpen
  \bibfield  {author} {\bibinfo {author} {\bibfnamefont {A.~A.}\ \bibnamefont
  {Starobinsky}},\ }\href {\doibase 10.1016/0370-2693(82)90541-X} {\bibfield
  {journal} {\bibinfo  {journal} {Phys. Lett. B}\ }\textbf {\bibinfo {volume}
  {117}},\ \bibinfo {pages} {175} (\bibinfo {year} {1982})}\BibitemShut
  {NoStop}%
\bibitem [{\citenamefont {Dimopoulos}\ \emph {et~al.}(2019)\citenamefont
  {Dimopoulos}, \citenamefont {Markkanen}, \citenamefont {Racioppi},\ and\
  \citenamefont {Vaskonen}}]{Dimopoulos:2019wew}%
  \BibitemOpen
  \bibfield  {author} {\bibinfo {author} {\bibfnamefont {K.}~\bibnamefont
  {Dimopoulos}}, \bibinfo {author} {\bibfnamefont {T.}~\bibnamefont
  {Markkanen}}, \bibinfo {author} {\bibfnamefont {A.}~\bibnamefont {Racioppi}},
  \ and\ \bibinfo {author} {\bibfnamefont {V.}~\bibnamefont {Vaskonen}},\
  }\href {\doibase 10.1088/1475-7516/2019/07/046} {\bibfield  {journal}
  {\bibinfo  {journal} {JCAP}\ }\textbf {\bibinfo {volume} {07}},\ \bibinfo
  {pages} {046} (\bibinfo {year} {2019})},\ \Eprint
  {http://arxiv.org/abs/1903.09598} {arXiv:1903.09598 [astro-ph.CO]}
  \BibitemShut {NoStop}%
\bibitem [{\citenamefont {Khlopov}\ and\ \citenamefont
  {Polnarev}(1980)}]{Khlopov:1980mg}%
  \BibitemOpen
  \bibfield  {author} {\bibinfo {author} {\bibfnamefont {M.~Y.}\ \bibnamefont
  {Khlopov}}\ and\ \bibinfo {author} {\bibfnamefont {A.~G.}\ \bibnamefont
  {Polnarev}},\ }\href {\doibase 10.1016/0370-2693(80)90624-3} {\bibfield
  {journal} {\bibinfo  {journal} {Phys. Lett. B}\ }\textbf {\bibinfo {volume}
  {97}},\ \bibinfo {pages} {383} (\bibinfo {year} {1980})}\BibitemShut
  {NoStop}%
\bibitem [{\citenamefont {Harada}\ \emph {et~al.}(2017)\citenamefont {Harada},
  \citenamefont {Yoo}, \citenamefont {Kohri},\ and\ \citenamefont
  {Nakao}}]{Harada:2017fjm}%
  \BibitemOpen
  \bibfield  {author} {\bibinfo {author} {\bibfnamefont {T.}~\bibnamefont
  {Harada}}, \bibinfo {author} {\bibfnamefont {C.-M.}\ \bibnamefont {Yoo}},
  \bibinfo {author} {\bibfnamefont {K.}~\bibnamefont {Kohri}}, \ and\ \bibinfo
  {author} {\bibfnamefont {K.-I.}\ \bibnamefont {Nakao}},\ }\href {\doibase
  10.1103/PhysRevD.96.083517} {\bibfield  {journal} {\bibinfo  {journal} {Phys.
  Rev. D}\ }\textbf {\bibinfo {volume} {96}},\ \bibinfo {pages} {083517}
  (\bibinfo {year} {2017})},\ \bibinfo {note} {[Erratum: Phys.Rev.D 99, 069904
  (2019)]},\ \Eprint {http://arxiv.org/abs/1707.03595} {arXiv:1707.03595
  [gr-qc]} \BibitemShut {NoStop}%
\bibitem [{\citenamefont {Carr}\ \emph {et~al.}(2017)\citenamefont {Carr},
  \citenamefont {Tenkanen},\ and\ \citenamefont {Vaskonen}}]{Carr:2017edp}%
  \BibitemOpen
  \bibfield  {author} {\bibinfo {author} {\bibfnamefont {B.}~\bibnamefont
  {Carr}}, \bibinfo {author} {\bibfnamefont {T.}~\bibnamefont {Tenkanen}}, \
  and\ \bibinfo {author} {\bibfnamefont {V.}~\bibnamefont {Vaskonen}},\ }\href
  {\doibase 10.1103/PhysRevD.96.063507} {\bibfield  {journal} {\bibinfo
  {journal} {Phys. Rev. D}\ }\textbf {\bibinfo {volume} {96}},\ \bibinfo
  {pages} {063507} (\bibinfo {year} {2017})},\ \Eprint
  {http://arxiv.org/abs/1706.03746} {arXiv:1706.03746 [astro-ph.CO]}
  \BibitemShut {NoStop}%
\bibitem [{\citenamefont {Carr}(1975)}]{Carr:1975qj}%
  \BibitemOpen
  \bibfield  {author} {\bibinfo {author} {\bibfnamefont {B.~J.}\ \bibnamefont
  {Carr}},\ }\href {\doibase 10.1086/153853} {\bibfield  {journal} {\bibinfo
  {journal} {Astrophys. J.}\ }\textbf {\bibinfo {volume} {201}},\ \bibinfo
  {pages} {1} (\bibinfo {year} {1975})}\BibitemShut {NoStop}%
\bibitem [{\citenamefont {Ellis}\ \emph
  {et~al.}(2019{\natexlab{b}})\citenamefont {Ellis}, \citenamefont {Lewicki},
  \citenamefont {No},\ and\ \citenamefont {Vaskonen}}]{Ellis:2019oqb}%
  \BibitemOpen
  \bibfield  {author} {\bibinfo {author} {\bibfnamefont {J.}~\bibnamefont
  {Ellis}}, \bibinfo {author} {\bibfnamefont {M.}~\bibnamefont {Lewicki}},
  \bibinfo {author} {\bibfnamefont {J.~M.}\ \bibnamefont {No}}, \ and\ \bibinfo
  {author} {\bibfnamefont {V.}~\bibnamefont {Vaskonen}},\ }\href {\doibase
  10.1088/1475-7516/2019/06/024} {\bibfield  {journal} {\bibinfo  {journal}
  {JCAP}\ }\textbf {\bibinfo {volume} {06}},\ \bibinfo {pages} {024} (\bibinfo
  {year} {2019}{\natexlab{b}})},\ \Eprint {http://arxiv.org/abs/1903.09642}
  {arXiv:1903.09642 [hep-ph]} \BibitemShut {NoStop}%
\bibitem [{\citenamefont {Lewicki}\ and\ \citenamefont
  {Vaskonen}(2020)}]{Lewicki:2020jiv}%
  \BibitemOpen
  \bibfield  {author} {\bibinfo {author} {\bibfnamefont {M.}~\bibnamefont
  {Lewicki}}\ and\ \bibinfo {author} {\bibfnamefont {V.}~\bibnamefont
  {Vaskonen}},\ }\href {\doibase 10.1140/epjc/s10052-020-08589-1} {\bibfield
  {journal} {\bibinfo  {journal} {Eur. Phys. J. C}\ }\textbf {\bibinfo {volume}
  {80}},\ \bibinfo {pages} {1003} (\bibinfo {year} {2020})},\ \Eprint
  {http://arxiv.org/abs/2007.04967} {arXiv:2007.04967 [astro-ph.CO]}
  \BibitemShut {NoStop}%
\bibitem [{\citenamefont {Lewicki}\ and\ \citenamefont
  {Vaskonen}(2021)}]{Lewicki:2020azd}%
  \BibitemOpen
  \bibfield  {author} {\bibinfo {author} {\bibfnamefont {M.}~\bibnamefont
  {Lewicki}}\ and\ \bibinfo {author} {\bibfnamefont {V.}~\bibnamefont
  {Vaskonen}},\ }\href {\doibase 10.1140/epjc/s10052-021-09232-3} {\bibfield
  {journal} {\bibinfo  {journal} {Eur. Phys. J. C}\ }\textbf {\bibinfo {volume}
  {81}},\ \bibinfo {pages} {437} (\bibinfo {year} {2021})},\ \Eprint
  {http://arxiv.org/abs/2012.07826} {arXiv:2012.07826 [astro-ph.CO]}
  \BibitemShut {NoStop}%
\bibitem [{\citenamefont {Saikawa}\ and\ \citenamefont
  {Shirai}(2018)}]{Saikawa:2018rcs}%
  \BibitemOpen
  \bibfield  {author} {\bibinfo {author} {\bibfnamefont {K.}~\bibnamefont
  {Saikawa}}\ and\ \bibinfo {author} {\bibfnamefont {S.}~\bibnamefont
  {Shirai}},\ }\href {\doibase 10.1088/1475-7516/2018/05/035} {\bibfield
  {journal} {\bibinfo  {journal} {JCAP}\ }\textbf {\bibinfo {volume} {05}},\
  \bibinfo {pages} {035} (\bibinfo {year} {2018})},\ \Eprint
  {http://arxiv.org/abs/1803.01038} {arXiv:1803.01038 [hep-ph]} \BibitemShut
  {NoStop}%
\bibitem [{\citenamefont {Thrane}\ and\ \citenamefont
  {Romano}(2013)}]{Thrane:2013oya}%
  \BibitemOpen
  \bibfield  {author} {\bibinfo {author} {\bibfnamefont {E.}~\bibnamefont
  {Thrane}}\ and\ \bibinfo {author} {\bibfnamefont {J.~D.}\ \bibnamefont
  {Romano}},\ }\href {\doibase 10.1103/PhysRevD.88.124032} {\bibfield
  {journal} {\bibinfo  {journal} {Phys. Rev. D}\ }\textbf {\bibinfo {volume}
  {88}},\ \bibinfo {pages} {124032} (\bibinfo {year} {2013})},\ \Eprint
  {http://arxiv.org/abs/1310.5300} {arXiv:1310.5300 [astro-ph.IM]} \BibitemShut
  {NoStop}%
\bibitem [{\citenamefont {Amaro-Seoane}\ \emph {et~al.}(2017)\citenamefont
  {Amaro-Seoane} \emph {et~al.}}]{Audley:2017drz}%
  \BibitemOpen
  \bibfield  {author} {\bibinfo {author} {\bibfnamefont {P.}~\bibnamefont
  {Amaro-Seoane}} \emph {et~al.} (\bibinfo {collaboration} {LISA}),\
  }\href@noop {} {\  (\bibinfo {year} {2017})},\ \Eprint
  {http://arxiv.org/abs/1702.00786} {arXiv:1702.00786 [astro-ph.IM]}
  \BibitemShut {NoStop}%
\bibitem [{\citenamefont {Punturo}\ \emph {et~al.}(2010)\citenamefont {Punturo}
  \emph {et~al.}}]{Punturo:2010zz}%
  \BibitemOpen
  \bibfield  {author} {\bibinfo {author} {\bibfnamefont {M.}~\bibnamefont
  {Punturo}} \emph {et~al.},\ }\href {\doibase 10.1088/0264-9381/27/19/194002}
  {\bibfield  {journal} {\bibinfo  {journal} {Class. Quant. Grav.}\ }\textbf
  {\bibinfo {volume} {27}},\ \bibinfo {pages} {194002} (\bibinfo {year}
  {2010})}\BibitemShut {NoStop}%
\bibitem [{\citenamefont {Hild}\ \emph {et~al.}(2011)\citenamefont {Hild} \emph
  {et~al.}}]{Hild:2010id}%
  \BibitemOpen
  \bibfield  {author} {\bibinfo {author} {\bibfnamefont {S.}~\bibnamefont
  {Hild}} \emph {et~al.},\ }\href {\doibase 10.1088/0264-9381/28/9/094013}
  {\bibfield  {journal} {\bibinfo  {journal} {Class. Quant. Grav.}\ }\textbf
  {\bibinfo {volume} {28}},\ \bibinfo {pages} {094013} (\bibinfo {year}
  {2011})},\ \Eprint {http://arxiv.org/abs/1012.0908} {arXiv:1012.0908 [gr-qc]}
  \BibitemShut {NoStop}%
\bibitem [{\citenamefont {El-Neaj}\ \emph {et~al.}(2020)\citenamefont {El-Neaj}
  \emph {et~al.}}]{Bertoldi:2019tck}%
  \BibitemOpen
  \bibfield  {author} {\bibinfo {author} {\bibfnamefont {Y.~A.}\ \bibnamefont
  {El-Neaj}} \emph {et~al.} (\bibinfo {collaboration} {AEDGE}),\ }\href
  {\doibase 10.1140/epjqt/s40507-020-0080-0} {\bibfield  {journal} {\bibinfo
  {journal} {EPJ Quant. Technol.}\ }\textbf {\bibinfo {volume} {7}},\ \bibinfo
  {pages} {6} (\bibinfo {year} {2020})},\ \Eprint
  {http://arxiv.org/abs/1908.00802} {arXiv:1908.00802 [gr-qc]} \BibitemShut
  {NoStop}%
\bibitem [{\citenamefont {Badurina}\ \emph {et~al.}(2020)\citenamefont
  {Badurina} \emph {et~al.}}]{Badurina:2019hst}%
  \BibitemOpen
  \bibfield  {author} {\bibinfo {author} {\bibfnamefont {L.}~\bibnamefont
  {Badurina}} \emph {et~al.},\ }\href {\doibase 10.1088/1475-7516/2020/05/011}
  {\bibfield  {journal} {\bibinfo  {journal} {JCAP}\ }\textbf {\bibinfo
  {volume} {05}},\ \bibinfo {pages} {011} (\bibinfo {year} {2020})},\ \Eprint
  {http://arxiv.org/abs/1911.11755} {arXiv:1911.11755 [astro-ph.CO]}
  \BibitemShut {NoStop}%
\bibitem [{\citenamefont {Graham}\ \emph {et~al.}(2016)\citenamefont {Graham},
  \citenamefont {Hogan}, \citenamefont {Kasevich},\ and\ \citenamefont
  {Rajendran}}]{Graham:2016plp}%
  \BibitemOpen
  \bibfield  {author} {\bibinfo {author} {\bibfnamefont {P.~W.}\ \bibnamefont
  {Graham}}, \bibinfo {author} {\bibfnamefont {J.~M.}\ \bibnamefont {Hogan}},
  \bibinfo {author} {\bibfnamefont {M.~A.}\ \bibnamefont {Kasevich}}, \ and\
  \bibinfo {author} {\bibfnamefont {S.}~\bibnamefont {Rajendran}},\ }\href
  {\doibase 10.1103/PhysRevD.94.104022} {\bibfield  {journal} {\bibinfo
  {journal} {Phys. Rev. D}\ }\textbf {\bibinfo {volume} {94}},\ \bibinfo
  {pages} {104022} (\bibinfo {year} {2016})},\ \Eprint
  {http://arxiv.org/abs/1606.01860} {arXiv:1606.01860 [physics.atom-ph]}
  \BibitemShut {NoStop}%
\bibitem [{\citenamefont {Graham}\ \emph {et~al.}(2017)\citenamefont {Graham},
  \citenamefont {Hogan}, \citenamefont {Kasevich}, \citenamefont {Rajendran},\
  and\ \citenamefont {Romani}}]{Graham:2017pmn}%
  \BibitemOpen
  \bibfield  {author} {\bibinfo {author} {\bibfnamefont {P.~W.}\ \bibnamefont
  {Graham}}, \bibinfo {author} {\bibfnamefont {J.~M.}\ \bibnamefont {Hogan}},
  \bibinfo {author} {\bibfnamefont {M.~A.}\ \bibnamefont {Kasevich}}, \bibinfo
  {author} {\bibfnamefont {S.}~\bibnamefont {Rajendran}}, \ and\ \bibinfo
  {author} {\bibfnamefont {R.~W.}\ \bibnamefont {Romani}} (\bibinfo
  {collaboration} {MAGIS}),\ }\href@noop {} {\  (\bibinfo {year} {2017})},\
  \Eprint {http://arxiv.org/abs/1711.02225} {arXiv:1711.02225 [astro-ph.IM]}
  \BibitemShut {NoStop}%
\bibitem [{\citenamefont {Janssen}\ \emph {et~al.}(2015)\citenamefont {Janssen}
  \emph {et~al.}}]{Janssen:2014dka}%
  \BibitemOpen
  \bibfield  {author} {\bibinfo {author} {\bibfnamefont {G.}~\bibnamefont
  {Janssen}} \emph {et~al.},\ }\href {\doibase 10.22323/1.215.0037} {\bibfield
  {journal} {\bibinfo  {journal} {PoS}\ }\textbf {\bibinfo {volume}
  {AASKA14}},\ \bibinfo {pages} {037} (\bibinfo {year} {2015})},\ \Eprint
  {http://arxiv.org/abs/1501.00127} {arXiv:1501.00127 [astro-ph.IM]}
  \BibitemShut {NoStop}%
\bibitem [{\citenamefont {Aasi}\ \emph {et~al.}(2015)\citenamefont {Aasi} \emph
  {et~al.}}]{TheLIGOScientific:2014jea}%
  \BibitemOpen
  \bibfield  {author} {\bibinfo {author} {\bibfnamefont {J.}~\bibnamefont
  {Aasi}} \emph {et~al.} (\bibinfo {collaboration} {LIGO Scientific}),\ }\href
  {\doibase 10.1088/0264-9381/32/7/074001} {\bibfield  {journal} {\bibinfo
  {journal} {Class. Quant. Grav.}\ }\textbf {\bibinfo {volume} {32}},\ \bibinfo
  {pages} {074001} (\bibinfo {year} {2015})},\ \Eprint
  {http://arxiv.org/abs/1411.4547} {arXiv:1411.4547 [gr-qc]} \BibitemShut
  {NoStop}%
\bibitem [{\citenamefont {Abbott}\ \emph {et~al.}(2019)\citenamefont {Abbott}
  \emph {et~al.}}]{LIGOScientific:2019vic}%
  \BibitemOpen
  \bibfield  {author} {\bibinfo {author} {\bibfnamefont {B.~P.}\ \bibnamefont
  {Abbott}} \emph {et~al.} (\bibinfo {collaboration} {LIGO Scientific,
  Virgo}),\ }\href {\doibase 10.1103/PhysRevD.100.061101} {\bibfield  {journal}
  {\bibinfo  {journal} {Phys. Rev. D}\ }\textbf {\bibinfo {volume} {100}},\
  \bibinfo {pages} {061101} (\bibinfo {year} {2019})},\ \Eprint
  {http://arxiv.org/abs/1903.02886} {arXiv:1903.02886 [gr-qc]} \BibitemShut
  {NoStop}%
\bibitem [{\citenamefont {Romero}\ \emph {et~al.}(2021)\citenamefont {Romero},
  \citenamefont {Martinovic}, \citenamefont {Callister}, \citenamefont {Guo},
  \citenamefont {Mart\'\i{}nez}, \citenamefont {Sakellariadou}, \citenamefont
  {Yang},\ and\ \citenamefont {Zhao}}]{Romero:2021kby}%
  \BibitemOpen
  \bibfield  {author} {\bibinfo {author} {\bibfnamefont {A.}~\bibnamefont
  {Romero}}, \bibinfo {author} {\bibfnamefont {K.}~\bibnamefont {Martinovic}},
  \bibinfo {author} {\bibfnamefont {T.~A.}\ \bibnamefont {Callister}}, \bibinfo
  {author} {\bibfnamefont {H.-K.}\ \bibnamefont {Guo}}, \bibinfo {author}
  {\bibfnamefont {M.}~\bibnamefont {Mart\'\i{}nez}}, \bibinfo {author}
  {\bibfnamefont {M.}~\bibnamefont {Sakellariadou}}, \bibinfo {author}
  {\bibfnamefont {F.-W.}\ \bibnamefont {Yang}}, \ and\ \bibinfo {author}
  {\bibfnamefont {Y.}~\bibnamefont {Zhao}},\ }\href {\doibase
  10.1103/PhysRevLett.126.151301} {\bibfield  {journal} {\bibinfo  {journal}
  {Phys. Rev. Lett.}\ }\textbf {\bibinfo {volume} {126}},\ \bibinfo {pages}
  {151301} (\bibinfo {year} {2021})},\ \Eprint
  {http://arxiv.org/abs/2102.01714} {arXiv:2102.01714 [hep-ph]} \BibitemShut
  {NoStop}%
\bibitem [{\citenamefont {Arzoumanian}\ \emph {et~al.}(2020)\citenamefont
  {Arzoumanian} \emph {et~al.}}]{Arzoumanian:2020vkk}%
  \BibitemOpen
  \bibfield  {author} {\bibinfo {author} {\bibfnamefont {Z.}~\bibnamefont
  {Arzoumanian}} \emph {et~al.} (\bibinfo {collaboration} {NANOGrav}),\ }\href
  {\doibase 10.3847/2041-8213/abd401} {\bibfield  {journal} {\bibinfo
  {journal} {Astrophys. J. Lett.}\ }\textbf {\bibinfo {volume} {905}},\
  \bibinfo {pages} {L34} (\bibinfo {year} {2020})},\ \Eprint
  {http://arxiv.org/abs/2009.04496} {arXiv:2009.04496 [astro-ph.HE]}
  \BibitemShut {NoStop}%
\bibitem [{\citenamefont {Ellis}\ and\ \citenamefont
  {Lewicki}(2021)}]{Ellis:2020ena}%
  \BibitemOpen
  \bibfield  {author} {\bibinfo {author} {\bibfnamefont {J.}~\bibnamefont
  {Ellis}}\ and\ \bibinfo {author} {\bibfnamefont {M.}~\bibnamefont
  {Lewicki}},\ }\href {\doibase 10.1103/PhysRevLett.126.041304} {\bibfield
  {journal} {\bibinfo  {journal} {Phys. Rev. Lett.}\ }\textbf {\bibinfo
  {volume} {126}},\ \bibinfo {pages} {041304} (\bibinfo {year} {2021})},\
  \Eprint {http://arxiv.org/abs/2009.06555} {arXiv:2009.06555 [astro-ph.CO]}
  \BibitemShut {NoStop}%
\bibitem [{\citenamefont {Vaskonen}\ and\ \citenamefont
  {Veerm\"ae}(2021)}]{Vaskonen:2020lbd}%
  \BibitemOpen
  \bibfield  {author} {\bibinfo {author} {\bibfnamefont {V.}~\bibnamefont
  {Vaskonen}}\ and\ \bibinfo {author} {\bibfnamefont {H.}~\bibnamefont
  {Veerm\"ae}},\ }\href {\doibase 10.1103/PhysRevLett.126.051303} {\bibfield
  {journal} {\bibinfo  {journal} {Phys. Rev. Lett.}\ }\textbf {\bibinfo
  {volume} {126}},\ \bibinfo {pages} {051303} (\bibinfo {year} {2021})},\
  \Eprint {http://arxiv.org/abs/2009.07832} {arXiv:2009.07832 [astro-ph.CO]}
  \BibitemShut {NoStop}%
\bibitem [{\citenamefont {De~Luca}\ \emph {et~al.}(2021)\citenamefont
  {De~Luca}, \citenamefont {Franciolini},\ and\ \citenamefont
  {Riotto}}]{DeLuca:2020agl}%
  \BibitemOpen
  \bibfield  {author} {\bibinfo {author} {\bibfnamefont {V.}~\bibnamefont
  {De~Luca}}, \bibinfo {author} {\bibfnamefont {G.}~\bibnamefont
  {Franciolini}}, \ and\ \bibinfo {author} {\bibfnamefont {A.}~\bibnamefont
  {Riotto}},\ }\href {\doibase 10.1103/PhysRevLett.126.041303} {\bibfield
  {journal} {\bibinfo  {journal} {Phys. Rev. Lett.}\ }\textbf {\bibinfo
  {volume} {126}},\ \bibinfo {pages} {041303} (\bibinfo {year} {2021})},\
  \Eprint {http://arxiv.org/abs/2009.08268} {arXiv:2009.08268 [astro-ph.CO]}
  \BibitemShut {NoStop}%
\bibitem [{\citenamefont {Kohri}\ and\ \citenamefont
  {Terada}(2021)}]{Kohri:2020qqd}%
  \BibitemOpen
  \bibfield  {author} {\bibinfo {author} {\bibfnamefont {K.}~\bibnamefont
  {Kohri}}\ and\ \bibinfo {author} {\bibfnamefont {T.}~\bibnamefont {Terada}},\
  }\href {\doibase 10.1016/j.physletb.2020.136040} {\bibfield  {journal}
  {\bibinfo  {journal} {Phys. Lett. B}\ }\textbf {\bibinfo {volume} {813}},\
  \bibinfo {pages} {136040} (\bibinfo {year} {2021})},\ \Eprint
  {http://arxiv.org/abs/2009.11853} {arXiv:2009.11853 [astro-ph.CO]}
  \BibitemShut {NoStop}%
\bibitem [{\citenamefont {Sugiyama}\ \emph {et~al.}(2021)\citenamefont
  {Sugiyama}, \citenamefont {Takhistov}, \citenamefont {Vitagliano},
  \citenamefont {Kusenko}, \citenamefont {Sasaki},\ and\ \citenamefont
  {Takada}}]{Sugiyama:2020roc}%
  \BibitemOpen
  \bibfield  {author} {\bibinfo {author} {\bibfnamefont {S.}~\bibnamefont
  {Sugiyama}}, \bibinfo {author} {\bibfnamefont {V.}~\bibnamefont {Takhistov}},
  \bibinfo {author} {\bibfnamefont {E.}~\bibnamefont {Vitagliano}}, \bibinfo
  {author} {\bibfnamefont {A.}~\bibnamefont {Kusenko}}, \bibinfo {author}
  {\bibfnamefont {M.}~\bibnamefont {Sasaki}}, \ and\ \bibinfo {author}
  {\bibfnamefont {M.}~\bibnamefont {Takada}},\ }\href {\doibase
  10.1016/j.physletb.2021.136097} {\bibfield  {journal} {\bibinfo  {journal}
  {Phys. Lett. B}\ }\textbf {\bibinfo {volume} {814}},\ \bibinfo {pages}
  {136097} (\bibinfo {year} {2021})},\ \Eprint
  {http://arxiv.org/abs/2010.02189} {arXiv:2010.02189 [astro-ph.CO]}
  \BibitemShut {NoStop}%
\bibitem [{\citenamefont {Dom\`enech}\ and\ \citenamefont
  {Pi}(2020)}]{Domenech:2020ers}%
  \BibitemOpen
  \bibfield  {author} {\bibinfo {author} {\bibfnamefont {G.}~\bibnamefont
  {Dom\`enech}}\ and\ \bibinfo {author} {\bibfnamefont {S.}~\bibnamefont
  {Pi}},\ }\href@noop {} {\  (\bibinfo {year} {2020})},\ \Eprint
  {http://arxiv.org/abs/2010.03976} {arXiv:2010.03976 [astro-ph.CO]}
  \BibitemShut {NoStop}%
\bibitem [{\citenamefont {Bhattacharya}\ \emph {et~al.}(2021)\citenamefont
  {Bhattacharya}, \citenamefont {Mohanty},\ and\ \citenamefont
  {Parashari}}]{Bhattacharya:2020lhc}%
  \BibitemOpen
  \bibfield  {author} {\bibinfo {author} {\bibfnamefont {S.}~\bibnamefont
  {Bhattacharya}}, \bibinfo {author} {\bibfnamefont {S.}~\bibnamefont
  {Mohanty}}, \ and\ \bibinfo {author} {\bibfnamefont {P.}~\bibnamefont
  {Parashari}},\ }\href {\doibase 10.1103/PhysRevD.103.063532} {\bibfield
  {journal} {\bibinfo  {journal} {Phys. Rev. D}\ }\textbf {\bibinfo {volume}
  {103}},\ \bibinfo {pages} {063532} (\bibinfo {year} {2021})},\ \Eprint
  {http://arxiv.org/abs/2010.05071} {arXiv:2010.05071 [astro-ph.CO]}
  \BibitemShut {NoStop}%
\bibitem [{\citenamefont {Blasi}\ \emph {et~al.}(2021)\citenamefont {Blasi},
  \citenamefont {Brdar},\ and\ \citenamefont {Schmitz}}]{Blasi:2020mfx}%
  \BibitemOpen
  \bibfield  {author} {\bibinfo {author} {\bibfnamefont {S.}~\bibnamefont
  {Blasi}}, \bibinfo {author} {\bibfnamefont {V.}~\bibnamefont {Brdar}}, \ and\
  \bibinfo {author} {\bibfnamefont {K.}~\bibnamefont {Schmitz}},\ }\href
  {\doibase 10.1103/PhysRevLett.126.041305} {\bibfield  {journal} {\bibinfo
  {journal} {Phys. Rev. Lett.}\ }\textbf {\bibinfo {volume} {126}},\ \bibinfo
  {pages} {041305} (\bibinfo {year} {2021})},\ \Eprint
  {http://arxiv.org/abs/2009.06607} {arXiv:2009.06607 [astro-ph.CO]}
  \BibitemShut {NoStop}%
\bibitem [{\citenamefont {Buchmuller}\ \emph {et~al.}(2020)\citenamefont
  {Buchmuller}, \citenamefont {Domcke},\ and\ \citenamefont
  {Schmitz}}]{Buchmuller:2020lbh}%
  \BibitemOpen
  \bibfield  {author} {\bibinfo {author} {\bibfnamefont {W.}~\bibnamefont
  {Buchmuller}}, \bibinfo {author} {\bibfnamefont {V.}~\bibnamefont {Domcke}},
  \ and\ \bibinfo {author} {\bibfnamefont {K.}~\bibnamefont {Schmitz}},\ }\href
  {\doibase 10.1016/j.physletb.2020.135914} {\bibfield  {journal} {\bibinfo
  {journal} {Phys. Lett. B}\ }\textbf {\bibinfo {volume} {811}},\ \bibinfo
  {pages} {135914} (\bibinfo {year} {2020})},\ \Eprint
  {http://arxiv.org/abs/2009.10649} {arXiv:2009.10649 [astro-ph.CO]}
  \BibitemShut {NoStop}%
\bibitem [{\citenamefont {Samanta}\ and\ \citenamefont
  {Datta}(2021)}]{Samanta:2020cdk}%
  \BibitemOpen
  \bibfield  {author} {\bibinfo {author} {\bibfnamefont {R.}~\bibnamefont
  {Samanta}}\ and\ \bibinfo {author} {\bibfnamefont {S.}~\bibnamefont
  {Datta}},\ }\href {\doibase 10.1007/JHEP05(2021)211} {\bibfield  {journal}
  {\bibinfo  {journal} {JHEP}\ }\textbf {\bibinfo {volume} {05}},\ \bibinfo
  {pages} {211} (\bibinfo {year} {2021})},\ \Eprint
  {http://arxiv.org/abs/2009.13452} {arXiv:2009.13452 [hep-ph]} \BibitemShut
  {NoStop}%
\bibitem [{\citenamefont {Chigusa}\ \emph {et~al.}(2020)\citenamefont
  {Chigusa}, \citenamefont {Nakai},\ and\ \citenamefont
  {Zheng}}]{Chigusa:2020rks}%
  \BibitemOpen
  \bibfield  {author} {\bibinfo {author} {\bibfnamefont {S.}~\bibnamefont
  {Chigusa}}, \bibinfo {author} {\bibfnamefont {Y.}~\bibnamefont {Nakai}}, \
  and\ \bibinfo {author} {\bibfnamefont {J.}~\bibnamefont {Zheng}},\
  }\href@noop {} {\  (\bibinfo {year} {2020})},\ \Eprint
  {http://arxiv.org/abs/2011.04090} {arXiv:2011.04090 [hep-ph]} \BibitemShut
  {NoStop}%
\bibitem [{\citenamefont {Ramberg}\ and\ \citenamefont
  {Visinelli}(2021)}]{Ramberg:2020oct}%
  \BibitemOpen
  \bibfield  {author} {\bibinfo {author} {\bibfnamefont {N.}~\bibnamefont
  {Ramberg}}\ and\ \bibinfo {author} {\bibfnamefont {L.}~\bibnamefont
  {Visinelli}},\ }\href {\doibase 10.1103/PhysRevD.103.063031} {\bibfield
  {journal} {\bibinfo  {journal} {Phys. Rev. D}\ }\textbf {\bibinfo {volume}
  {103}},\ \bibinfo {pages} {063031} (\bibinfo {year} {2021})},\ \Eprint
  {http://arxiv.org/abs/2012.06882} {arXiv:2012.06882 [astro-ph.CO]}
  \BibitemShut {NoStop}%
\bibitem [{\citenamefont {Nakai}\ \emph {et~al.}(2021)\citenamefont {Nakai},
  \citenamefont {Suzuki}, \citenamefont {Takahashi},\ and\ \citenamefont
  {Yamada}}]{Nakai:2020oit}%
  \BibitemOpen
  \bibfield  {author} {\bibinfo {author} {\bibfnamefont {Y.}~\bibnamefont
  {Nakai}}, \bibinfo {author} {\bibfnamefont {M.}~\bibnamefont {Suzuki}},
  \bibinfo {author} {\bibfnamefont {F.}~\bibnamefont {Takahashi}}, \ and\
  \bibinfo {author} {\bibfnamefont {M.}~\bibnamefont {Yamada}},\ }\href
  {\doibase 10.1016/j.physletb.2021.136238} {\bibfield  {journal} {\bibinfo
  {journal} {Phys. Lett. B}\ }\textbf {\bibinfo {volume} {816}},\ \bibinfo
  {pages} {136238} (\bibinfo {year} {2021})},\ \Eprint
  {http://arxiv.org/abs/2009.09754} {arXiv:2009.09754 [astro-ph.CO]}
  \BibitemShut {NoStop}%
\bibitem [{\citenamefont {Neronov}\ \emph {et~al.}(2021)\citenamefont
  {Neronov}, \citenamefont {Roper~Pol}, \citenamefont {Caprini},\ and\
  \citenamefont {Semikoz}}]{Neronov:2020qrl}%
  \BibitemOpen
  \bibfield  {author} {\bibinfo {author} {\bibfnamefont {A.}~\bibnamefont
  {Neronov}}, \bibinfo {author} {\bibfnamefont {A.}~\bibnamefont {Roper~Pol}},
  \bibinfo {author} {\bibfnamefont {C.}~\bibnamefont {Caprini}}, \ and\
  \bibinfo {author} {\bibfnamefont {D.}~\bibnamefont {Semikoz}},\ }\href
  {\doibase 10.1103/PhysRevD.103.L041302} {\bibfield  {journal} {\bibinfo
  {journal} {Phys. Rev. D}\ }\textbf {\bibinfo {volume} {103}},\ \bibinfo
  {pages} {041302} (\bibinfo {year} {2021})},\ \Eprint
  {http://arxiv.org/abs/2009.14174} {arXiv:2009.14174 [astro-ph.CO]}
  \BibitemShut {NoStop}%
\bibitem [{\citenamefont {Vagnozzi}(2021)}]{Vagnozzi:2020gtf}%
  \BibitemOpen
  \bibfield  {author} {\bibinfo {author} {\bibfnamefont {S.}~\bibnamefont
  {Vagnozzi}},\ }\href {\doibase 10.1093/mnrasl/slaa203} {\bibfield  {journal}
  {\bibinfo  {journal} {Mon. Not. Roy. Astron. Soc.}\ }\textbf {\bibinfo
  {volume} {502}},\ \bibinfo {pages} {L11} (\bibinfo {year} {2021})},\ \Eprint
  {http://arxiv.org/abs/2009.13432} {arXiv:2009.13432 [astro-ph.CO]}
  \BibitemShut {NoStop}%
\bibitem [{\citenamefont {Li}\ \emph {et~al.}(2021)\citenamefont {Li},
  \citenamefont {Ye},\ and\ \citenamefont {Piao}}]{Li:2020cjj}%
  \BibitemOpen
  \bibfield  {author} {\bibinfo {author} {\bibfnamefont {H.-H.}\ \bibnamefont
  {Li}}, \bibinfo {author} {\bibfnamefont {G.}~\bibnamefont {Ye}}, \ and\
  \bibinfo {author} {\bibfnamefont {Y.-S.}\ \bibnamefont {Piao}},\ }\href
  {\doibase 10.1016/j.physletb.2021.136211} {\bibfield  {journal} {\bibinfo
  {journal} {Phys. Lett. B}\ }\textbf {\bibinfo {volume} {816}},\ \bibinfo
  {pages} {136211} (\bibinfo {year} {2021})},\ \Eprint
  {http://arxiv.org/abs/2009.14663} {arXiv:2009.14663 [astro-ph.CO]}
  \BibitemShut {NoStop}%
\bibitem [{\citenamefont {Kuroyanagi}\ \emph {et~al.}(2021)\citenamefont
  {Kuroyanagi}, \citenamefont {Takahashi},\ and\ \citenamefont
  {Yokoyama}}]{Kuroyanagi:2020sfw}%
  \BibitemOpen
  \bibfield  {author} {\bibinfo {author} {\bibfnamefont {S.}~\bibnamefont
  {Kuroyanagi}}, \bibinfo {author} {\bibfnamefont {T.}~\bibnamefont
  {Takahashi}}, \ and\ \bibinfo {author} {\bibfnamefont {S.}~\bibnamefont
  {Yokoyama}},\ }\href {\doibase 10.1088/1475-7516/2021/01/071} {\bibfield
  {journal} {\bibinfo  {journal} {JCAP}\ }\textbf {\bibinfo {volume} {01}},\
  \bibinfo {pages} {071} (\bibinfo {year} {2021})},\ \Eprint
  {http://arxiv.org/abs/2011.03323} {arXiv:2011.03323 [astro-ph.CO]}
  \BibitemShut {NoStop}%
\bibitem [{\citenamefont {Liu}\ \emph {et~al.}(2021)\citenamefont {Liu},
  \citenamefont {Cai},\ and\ \citenamefont {Guo}}]{Liu:2020mru}%
  \BibitemOpen
  \bibfield  {author} {\bibinfo {author} {\bibfnamefont {J.}~\bibnamefont
  {Liu}}, \bibinfo {author} {\bibfnamefont {R.-G.}\ \bibnamefont {Cai}}, \ and\
  \bibinfo {author} {\bibfnamefont {Z.-K.}\ \bibnamefont {Guo}},\ }\href
  {\doibase 10.1103/PhysRevLett.126.141303} {\bibfield  {journal} {\bibinfo
  {journal} {Phys. Rev. Lett.}\ }\textbf {\bibinfo {volume} {126}},\ \bibinfo
  {pages} {141303} (\bibinfo {year} {2021})},\ \Eprint
  {http://arxiv.org/abs/2010.03225} {arXiv:2010.03225 [astro-ph.CO]}
  \BibitemShut {NoStop}%
\bibitem [{\citenamefont {Chiang}\ and\ \citenamefont
  {Lu}(2021)}]{Chiang:2020aui}%
  \BibitemOpen
  \bibfield  {author} {\bibinfo {author} {\bibfnamefont {C.-W.}\ \bibnamefont
  {Chiang}}\ and\ \bibinfo {author} {\bibfnamefont {B.-Q.}\ \bibnamefont
  {Lu}},\ }\href {\doibase 10.1088/1475-7516/2021/05/049} {\bibfield  {journal}
  {\bibinfo  {journal} {JCAP}\ }\textbf {\bibinfo {volume} {05}},\ \bibinfo
  {pages} {049} (\bibinfo {year} {2021})},\ \Eprint
  {http://arxiv.org/abs/2012.14071} {arXiv:2012.14071 [hep-ph]} \BibitemShut
  {NoStop}%
\bibitem [{\citenamefont {Ratzinger}\ and\ \citenamefont
  {Schwaller}(2021)}]{Ratzinger:2020koh}%
  \BibitemOpen
  \bibfield  {author} {\bibinfo {author} {\bibfnamefont {W.}~\bibnamefont
  {Ratzinger}}\ and\ \bibinfo {author} {\bibfnamefont {P.}~\bibnamefont
  {Schwaller}},\ }\href {\doibase 10.21468/SciPostPhys.10.2.047} {\bibfield
  {journal} {\bibinfo  {journal} {SciPost Phys.}\ }\textbf {\bibinfo {volume}
  {10}},\ \bibinfo {pages} {047} (\bibinfo {year} {2021})},\ \Eprint
  {http://arxiv.org/abs/2009.11875} {arXiv:2009.11875 [astro-ph.CO]}
  \BibitemShut {NoStop}%
\bibitem [{\citenamefont {Namba}\ and\ \citenamefont
  {Suzuki}(2020)}]{Namba:2020kij}%
  \BibitemOpen
  \bibfield  {author} {\bibinfo {author} {\bibfnamefont {R.}~\bibnamefont
  {Namba}}\ and\ \bibinfo {author} {\bibfnamefont {M.}~\bibnamefont {Suzuki}},\
  }\href {\doibase 10.1103/PhysRevD.102.123527} {\bibfield  {journal} {\bibinfo
   {journal} {Phys. Rev. D}\ }\textbf {\bibinfo {volume} {102}},\ \bibinfo
  {pages} {123527} (\bibinfo {year} {2020})},\ \Eprint
  {http://arxiv.org/abs/2009.13909} {arXiv:2009.13909 [astro-ph.CO]}
  \BibitemShut {NoStop}%
\bibitem [{\citenamefont {Tahara}\ and\ \citenamefont
  {Kobayashi}(2020)}]{Tahara:2020fmn}%
  \BibitemOpen
  \bibfield  {author} {\bibinfo {author} {\bibfnamefont {H.~W.~H.}\
  \bibnamefont {Tahara}}\ and\ \bibinfo {author} {\bibfnamefont
  {T.}~\bibnamefont {Kobayashi}},\ }\href {\doibase
  10.1103/PhysRevD.102.123533} {\bibfield  {journal} {\bibinfo  {journal}
  {Phys. Rev. D}\ }\textbf {\bibinfo {volume} {102}},\ \bibinfo {pages}
  {123533} (\bibinfo {year} {2020})},\ \Eprint
  {http://arxiv.org/abs/2011.01605} {arXiv:2011.01605 [gr-qc]} \BibitemShut
  {NoStop}%
\bibitem [{\citenamefont {Cai}\ and\ \citenamefont {Piao}(2021)}]{Cai:2020qpu}%
  \BibitemOpen
  \bibfield  {author} {\bibinfo {author} {\bibfnamefont {Y.}~\bibnamefont
  {Cai}}\ and\ \bibinfo {author} {\bibfnamefont {Y.-S.}\ \bibnamefont {Piao}},\
  }\href {\doibase 10.1103/PhysRevD.103.083521} {\bibfield  {journal} {\bibinfo
   {journal} {Phys. Rev. D}\ }\textbf {\bibinfo {volume} {103}},\ \bibinfo
  {pages} {083521} (\bibinfo {year} {2021})},\ \Eprint
  {http://arxiv.org/abs/2012.11304} {arXiv:2012.11304 [gr-qc]} \BibitemShut
  {NoStop}%
\bibitem [{\citenamefont {Bian}\ \emph {et~al.}(2021)\citenamefont {Bian},
  \citenamefont {Cai}, \citenamefont {Liu}, \citenamefont {Yang},\ and\
  \citenamefont {Zhou}}]{Bian:2021lmz}%
  \BibitemOpen
  \bibfield  {author} {\bibinfo {author} {\bibfnamefont {L.}~\bibnamefont
  {Bian}}, \bibinfo {author} {\bibfnamefont {R.-G.}\ \bibnamefont {Cai}},
  \bibinfo {author} {\bibfnamefont {J.}~\bibnamefont {Liu}}, \bibinfo {author}
  {\bibfnamefont {X.-Y.}\ \bibnamefont {Yang}}, \ and\ \bibinfo {author}
  {\bibfnamefont {R.}~\bibnamefont {Zhou}},\ }\href {\doibase
  10.1103/PhysRevD.103.L081301} {\bibfield  {journal} {\bibinfo  {journal}
  {Phys. Rev. D}\ }\textbf {\bibinfo {volume} {103}},\ \bibinfo {pages}
  {L081301} (\bibinfo {year} {2021})},\ \Eprint
  {http://arxiv.org/abs/2009.13893} {arXiv:2009.13893 [astro-ph.CO]}
  \BibitemShut {NoStop}%
\bibitem [{\citenamefont {Blanco-Pillado}\ \emph {et~al.}(2021)\citenamefont
  {Blanco-Pillado}, \citenamefont {Olum},\ and\ \citenamefont
  {Wachter}}]{Blanco-Pillado:2021ygr}%
  \BibitemOpen
  \bibfield  {author} {\bibinfo {author} {\bibfnamefont {J.~J.}\ \bibnamefont
  {Blanco-Pillado}}, \bibinfo {author} {\bibfnamefont {K.~D.}\ \bibnamefont
  {Olum}}, \ and\ \bibinfo {author} {\bibfnamefont {J.~M.}\ \bibnamefont
  {Wachter}},\ }\href {\doibase 10.1103/PhysRevD.103.103512} {\bibfield
  {journal} {\bibinfo  {journal} {Phys. Rev. D}\ }\textbf {\bibinfo {volume}
  {103}},\ \bibinfo {pages} {103512} (\bibinfo {year} {2021})},\ \Eprint
  {http://arxiv.org/abs/2102.08194} {arXiv:2102.08194 [astro-ph.CO]}
  \BibitemShut {NoStop}%
\bibitem [{\citenamefont {Brandenburg}\ \emph {et~al.}(2021)\citenamefont
  {Brandenburg}, \citenamefont {Clarke}, \citenamefont {He},\ and\
  \citenamefont {Kahniashvili}}]{Brandenburg:2021tmp}%
  \BibitemOpen
  \bibfield  {author} {\bibinfo {author} {\bibfnamefont {A.}~\bibnamefont
  {Brandenburg}}, \bibinfo {author} {\bibfnamefont {E.}~\bibnamefont {Clarke}},
  \bibinfo {author} {\bibfnamefont {Y.}~\bibnamefont {He}}, \ and\ \bibinfo
  {author} {\bibfnamefont {T.}~\bibnamefont {Kahniashvili}},\ }\href@noop {} {\
   (\bibinfo {year} {2021})},\ \Eprint {http://arxiv.org/abs/2102.12428}
  {arXiv:2102.12428 [astro-ph.CO]} \BibitemShut {NoStop}%
\bibitem [{\citenamefont {Arzoumanian}\ \emph {et~al.}(2021)\citenamefont
  {Arzoumanian} \emph {et~al.}}]{Arzoumanian:2021teu}%
  \BibitemOpen
  \bibfield  {author} {\bibinfo {author} {\bibfnamefont {Z.}~\bibnamefont
  {Arzoumanian}} \emph {et~al.} (\bibinfo {collaboration} {NANOGrav}),\
  }\href@noop {} {\  (\bibinfo {year} {2021})},\ \Eprint
  {http://arxiv.org/abs/2104.13930} {arXiv:2104.13930 [astro-ph.CO]}
  \BibitemShut {NoStop}%
\bibitem [{\citenamefont {Matarrese}\ \emph {et~al.}(1994)\citenamefont
  {Matarrese}, \citenamefont {Pantano},\ and\ \citenamefont
  {Saez}}]{Matarrese:1993zf}%
  \BibitemOpen
  \bibfield  {author} {\bibinfo {author} {\bibfnamefont {S.}~\bibnamefont
  {Matarrese}}, \bibinfo {author} {\bibfnamefont {O.}~\bibnamefont {Pantano}},
  \ and\ \bibinfo {author} {\bibfnamefont {D.}~\bibnamefont {Saez}},\ }\href
  {\doibase 10.1103/PhysRevLett.72.320} {\bibfield  {journal} {\bibinfo
  {journal} {Phys. Rev. Lett.}\ }\textbf {\bibinfo {volume} {72}},\ \bibinfo
  {pages} {320} (\bibinfo {year} {1994})},\ \Eprint
  {http://arxiv.org/abs/astro-ph/9310036} {arXiv:astro-ph/9310036} \BibitemShut
  {NoStop}%
\bibitem [{\citenamefont {Matarrese}\ \emph {et~al.}(1998)\citenamefont
  {Matarrese}, \citenamefont {Mollerach},\ and\ \citenamefont
  {Bruni}}]{Matarrese:1997ay}%
  \BibitemOpen
  \bibfield  {author} {\bibinfo {author} {\bibfnamefont {S.}~\bibnamefont
  {Matarrese}}, \bibinfo {author} {\bibfnamefont {S.}~\bibnamefont
  {Mollerach}}, \ and\ \bibinfo {author} {\bibfnamefont {M.}~\bibnamefont
  {Bruni}},\ }\href {\doibase 10.1103/PhysRevD.58.043504} {\bibfield  {journal}
  {\bibinfo  {journal} {Phys. Rev. D}\ }\textbf {\bibinfo {volume} {58}},\
  \bibinfo {pages} {043504} (\bibinfo {year} {1998})},\ \Eprint
  {http://arxiv.org/abs/astro-ph/9707278} {arXiv:astro-ph/9707278} \BibitemShut
  {NoStop}%
\bibitem [{\citenamefont {Nakamura}(2007)}]{Nakamura:2004rm}%
  \BibitemOpen
  \bibfield  {author} {\bibinfo {author} {\bibfnamefont {K.}~\bibnamefont
  {Nakamura}},\ }\href {\doibase 10.1143/PTP.117.17} {\bibfield  {journal}
  {\bibinfo  {journal} {Prog. Theor. Phys.}\ }\textbf {\bibinfo {volume}
  {117}},\ \bibinfo {pages} {17} (\bibinfo {year} {2007})},\ \Eprint
  {http://arxiv.org/abs/gr-qc/0605108} {arXiv:gr-qc/0605108} \BibitemShut
  {NoStop}%
\bibitem [{\citenamefont {Ananda}\ \emph {et~al.}(2007)\citenamefont {Ananda},
  \citenamefont {Clarkson},\ and\ \citenamefont {Wands}}]{Ananda:2006af}%
  \BibitemOpen
  \bibfield  {author} {\bibinfo {author} {\bibfnamefont {K.~N.}\ \bibnamefont
  {Ananda}}, \bibinfo {author} {\bibfnamefont {C.}~\bibnamefont {Clarkson}}, \
  and\ \bibinfo {author} {\bibfnamefont {D.}~\bibnamefont {Wands}},\ }\href
  {\doibase 10.1103/PhysRevD.75.123518} {\bibfield  {journal} {\bibinfo
  {journal} {Phys. Rev. D}\ }\textbf {\bibinfo {volume} {75}},\ \bibinfo
  {pages} {123518} (\bibinfo {year} {2007})},\ \Eprint
  {http://arxiv.org/abs/gr-qc/0612013} {arXiv:gr-qc/0612013} \BibitemShut
  {NoStop}%
\bibitem [{\citenamefont {Baumann}\ \emph {et~al.}(2007)\citenamefont
  {Baumann}, \citenamefont {Steinhardt}, \citenamefont {Takahashi},\ and\
  \citenamefont {Ichiki}}]{Baumann:2007zm}%
  \BibitemOpen
  \bibfield  {author} {\bibinfo {author} {\bibfnamefont {D.}~\bibnamefont
  {Baumann}}, \bibinfo {author} {\bibfnamefont {P.~J.}\ \bibnamefont
  {Steinhardt}}, \bibinfo {author} {\bibfnamefont {K.}~\bibnamefont
  {Takahashi}}, \ and\ \bibinfo {author} {\bibfnamefont {K.}~\bibnamefont
  {Ichiki}},\ }\href {\doibase 10.1103/PhysRevD.76.084019} {\bibfield
  {journal} {\bibinfo  {journal} {Phys. Rev. D}\ }\textbf {\bibinfo {volume}
  {76}},\ \bibinfo {pages} {084019} (\bibinfo {year} {2007})},\ \Eprint
  {http://arxiv.org/abs/hep-th/0703290} {arXiv:hep-th/0703290} \BibitemShut
  {NoStop}%
\bibitem [{\citenamefont {Kohri}\ and\ \citenamefont
  {Terada}(2018)}]{Kohri:2018awv}%
  \BibitemOpen
  \bibfield  {author} {\bibinfo {author} {\bibfnamefont {K.}~\bibnamefont
  {Kohri}}\ and\ \bibinfo {author} {\bibfnamefont {T.}~\bibnamefont {Terada}},\
  }\href {\doibase 10.1103/PhysRevD.97.123532} {\bibfield  {journal} {\bibinfo
  {journal} {Phys. Rev. D}\ }\textbf {\bibinfo {volume} {97}},\ \bibinfo
  {pages} {123532} (\bibinfo {year} {2018})},\ \Eprint
  {http://arxiv.org/abs/1804.08577} {arXiv:1804.08577 [gr-qc]} \BibitemShut
  {NoStop}%
\bibitem [{\citenamefont {Espinosa}\ \emph {et~al.}(2018)\citenamefont
  {Espinosa}, \citenamefont {Racco},\ and\ \citenamefont
  {Riotto}}]{Espinosa:2018eve}%
  \BibitemOpen
  \bibfield  {author} {\bibinfo {author} {\bibfnamefont {J.~R.}\ \bibnamefont
  {Espinosa}}, \bibinfo {author} {\bibfnamefont {D.}~\bibnamefont {Racco}}, \
  and\ \bibinfo {author} {\bibfnamefont {A.}~\bibnamefont {Riotto}},\ }\href
  {\doibase 10.1088/1475-7516/2018/09/012} {\bibfield  {journal} {\bibinfo
  {journal} {JCAP}\ }\textbf {\bibinfo {volume} {09}},\ \bibinfo {pages} {012}
  (\bibinfo {year} {2018})},\ \Eprint {http://arxiv.org/abs/1804.07732}
  {arXiv:1804.07732 [hep-ph]} \BibitemShut {NoStop}%
\bibitem [{\citenamefont {Inomata}\ and\ \citenamefont
  {Terada}(2020)}]{Inomata:2019yww}%
  \BibitemOpen
  \bibfield  {author} {\bibinfo {author} {\bibfnamefont {K.}~\bibnamefont
  {Inomata}}\ and\ \bibinfo {author} {\bibfnamefont {T.}~\bibnamefont
  {Terada}},\ }\href {\doibase 10.1103/PhysRevD.101.023523} {\bibfield
  {journal} {\bibinfo  {journal} {Phys. Rev. D}\ }\textbf {\bibinfo {volume}
  {101}},\ \bibinfo {pages} {023523} (\bibinfo {year} {2020})},\ \Eprint
  {http://arxiv.org/abs/1912.00785} {arXiv:1912.00785 [gr-qc]} \BibitemShut
  {NoStop}%
\end{thebibliography}%
\end{document}